\documentclass[a4paper,11pt]{article}
\pdfoutput=1

\usepackage[utf8]{inputenc}
\usepackage{a4wide}
\usepackage{graphicx}
\graphicspath{{figures/}} 
\usepackage[small,bf]{caption}
\usepackage{amssymb}
\usepackage{amsmath}
\usepackage{slashed}
\usepackage{mathtools}
\usepackage{cite} 
\usepackage{hyperref}
\usepackage{xcolor}

\newcommand{\be}{\begin{equation}}
\newcommand{\ee}{\end{equation}}
\newcommand{\ben}{\begin{eqnarray}}
\newcommand{\een}{\end{eqnarray}}
\newcommand{\bi}{\begin{itemize}}
\newcommand{\ei}{\end{itemize}}
\newcommand\barparenb[1]{\overset{%
   \scalebox{0.4}{$(\mkern-1mu-\mkern-1mu)$}}{#1}}

\numberwithin{equation}{section}
\makeatletter
\g@addto@macro\bfseries{\boldmath}
\makeatother

\begin{document}

\begin{titlepage}
\begin{flushright}
FTUAM-19-24 \\
IFT-UAM/CSIC-19-162
\end{flushright}
\vspace*{0.8cm}

\begin{center}
{\Large\bf Physics potential of the ESS$\nu$SB}\\[0.8cm]
M.~Blennow,$^{a,b}$
E.~Fernandez-Martinez,$^a$
T.~Ota$^a$
and S.~Rosauro-Alcaraz$^a$\\[0.4cm]
$^{a}$\,{\it Departamento de F\'isica Te\'orica and Instituto de F\'{\i}sica Te\'orica, IFT-UAM/CSIC,\\
Universidad Aut\'onoma de Madrid, Cantoblanco, 28049, Madrid, Spain} \\
$^{b}$\,{\it Department of Physics, School of Engineering Sciences, \\ KTH Royal Institute of Technology, AlbaNova University Center, \\ Roslagstullsbacken 21, SE--106 91 Stockholm, Sweden}\\
\end{center}
\vspace{0.8cm}

\begin{abstract}
The ESS$\nu$SB project proposes to base a neutrino ``Super Beam'' of unprecedented luminosity at the European Spallation Source. The original proposal identified the second peak of the oscillation probability as the optimal to maximize the discovery potential to leptonic CP violation. However this choice reduces the statistics at the detector and penalizes other complementary searches such as the determination of the atmospheric oscillation parameters, particularly the octant of $\theta_{23}$ as well as the neutrino mass ordering. We explore how these shortcomings can be alleviated by the combination of the beam data with the atmospheric neutrino sample that would also be collected at the detector. We find that the combination not only improves very significantly these drawbacks, but also enhances both the CP violation discovery potential and the precision in the measurement of the CP violating phase, for which the facility was originally optimized, by lifting parametric degeneracies. We then reassess the optimization of the ESS$\nu$SB setup when the atmospheric neutrino sample is considered, with an emphasis in performing a measurement of the CP violating phase as precise as possible. We find that for the presently preferred value of $\delta \sim -\pi/2$, shorter baselines and longer running time in neutrino mode would be optimal. In these conditions, a measurement better than $14^\circ$ would be achievable for any value of the $\theta_{23}$ octant and the mass ordering. Conversely, if present and next generation facilities were not able to discover CP violation, longer baselines and more even splitting between neutrino and neutrino modes would be preferable. These choices would allow a $5 \sigma$ discovery of CP violation for around a $60\%$ of the possible values of $\delta$ and to determine its value with a precision around $6^\circ$ if it is close to $0$ or $\pi$.

\end{abstract}
\end{titlepage}
\section{Introduction}
\label{sec:introduction}

After the discovery of a non-zero $\theta_{13}$~\cite{An:2012eh,Ahn:2012nd,Abe:2012tg,Adamson:2011qu,Abe:2011sj} the emerging picture from the last decades of neutrino oscillation searches consolidates a structure for the PMNS matrix~\cite{Pontecorvo:1957cp,Pontecorvo:1957qd,Maki:1960ut,Maki:1962mu,Pontecorvo:1967fh} describing lepton flavour mixing strikingly different from its CKM counterpart in the quark sector, making the Standard Model flavour puzzle even more intriguing. Far from the hierarchical structure described through the tiny mixing angles of the CKM, large mixing angles characterize the lepton mixing. The ``atmospheric'' mixing angle $\theta_{23}$ is presently compatible with maximal mixing as well as with a large but non-maximal value in either the first or the second octant. Similarly, the ``solar'' mixing angle $\theta_{12}$ is around $33^\circ$ and only $\theta_{13} \sim 8-9^\circ$ is relatively small and its value is still comparable in magnitude to the Cabibbo angle, the largest in the CKM. The large mixing opens the window to the present and next generation of neutrino oscillation experiments to tackle new questions that could provide answers to fundamental open problems. 

Present experiments such as T2K~\cite{Abe:2017uxa, Abe:2019vii} and NO$\nu$A~\cite{Acero:2019ksn} have started to provide the first hints on the potentially CP-violating phase $\delta$. The discovery of the violation of the particle-antiparticle symmetry in the lepton sector would be extremely suggestive, given that CP-violation is a necessary ingredient to explain the matter over antimatter excess to which we owe our existence and that the CKM contribution has been shown to be insufficient~\cite{Gavela:1993ts,Gavela:1994dt} for this purpose. Similarly, present neutrino oscillation experiments already show some preference for normal ordering (positive $\Delta m^2_{31}$) with respect to inverted ordering. This parameter is a fundamental input to combine with the searches for the neutrinoless double beta decay process in order to probe the Majorana nature of neutrinos. Finally, present experiments as well as their successors T2HK~\cite{Abe:2015zbg} and DUNE~\cite{Acciarri:2015uup} will also provide even more precise measurements of the oscillation parameters that could hold the key to discriminate among different flavour models addressing the flavour puzzle.    

The European Spallation Source (ESS) at Lund provides an opportunity to build a new-generation, long-baseline neutrino oscillation experiment with an unprecedented neutrino luminosity through an upgrade of the ESS Linac~\cite{Baussan:2013zcy}. Its $2.5$~GeV protons would lead to a rather low energy neutrino flux, between 200 and 600~MeV. This energy range is very well suited for a water Cerenkov detector of the MEMPHYS type~\cite{deBellefon:2006vq,Agostino:2012fd}. In Ref.~\cite{Baussan:2013zcy} a greenfield study optimizing the physics reach to leptonic CP-violation was performed for this ESS neutrino Super-Beam facility (ESS$\nu$SB). Interestingly, the outcome of this optimization, as well as follow-up studies~\cite{Agarwalla:2014tpa,Chakraborty:2017ccm,Chakraborty:2019jlv}, was that the best baseline at which to study the neutrino beam from the ESS facility at a MEMPHYS-type detector would be between 400 and 600~km. Two candidate mines that could host the detector were identified: Garpenberg at 540~km and Zinkgruvan at 360~km from the ESS site. This choice makes the ESS$\nu$SB design unique, as the neutrino flux observed by the detector mainly corresponds to the second maximum of the $\nu_\mu \to \nu_e$ oscillation probability, with a marginal contribution of events at the first oscillation peak.

For the value of $\theta_{13} = 8.6^\circ$ currently preferred~\cite{Esteban:2018azc} by Daya Bay~\cite{Adey:2018zwh} and RENO~\cite{Bak:2018ydk}, the ``atmospheric'' term of the $\nu_\mu \to \nu_e$ oscillation probability~\cite{Cervera:2000kp}, which is governed by oscillations driven by the large frequency $\Delta m^2_{31}$ and with an amplitude $\sin^2 2\theta_{13}$, dominates over the sub-leading ``solar'' term driven by $\Delta m^2_{21}$ with amplitude $\sin^2 2\theta_{12}$ at the first oscillation maximum. Thus, the interference between the two, which is the only term dependent on the yet unknown CP-violating phase $\delta$, will also be a sub-leading contribution to the full oscillation probability at the first peak and potentially hidden by systematic uncertainties. Conversely, at the second oscillation maximum the slower ``solar'' oscillation has had more time to develop and thus the CP-violating interference term can give a significant contribution to the oscillation probability, thus increasing the sensitivity to CP violation~\cite{Coloma:2011pg}. 

The price to pay in order to observe the oscillation probability at its second maximum is high. Despite this being the optimal choice to maximize the dependence of the oscillation probability on the leptonic CP violating phase, the ratio of the oscillation baseline to the neutrino energy ($L/E$) needs to be a factor 3 larger compared to the first maximum. This implies roughly an order of magnitude less statistics than if the experiment had been designed at the first peak. Indeed, the neutrino flux decreases with $L^{-2}$ from the beam divergence and the neutrino cross section and beam collimation increase with the neutrino energy. Despite the unprecedented neutrino luminosity from the upgraded ESS linac and the megaton-class MEMPHYS detector, only around 100 signal events for each beam polarity would be accumulated after 10 years data taking (2 years in neutrinos and 8 years in antineutrinos) at the 540~km Garpenberg baseline (see Fig.~7 of Ref.~\cite{Baussan:2013zcy}). Conversely, the 360~km Zinkgruvan baseline has a 2.25 times larger neutrino flux. However, the neutrino spectrum for this baseline is rather centered at the first oscillation minimum while the first and second peaks are sampled by the high and low energy tails respectively. Overall this gives similar statistics at the second oscillation maximum when compared to the Garpenberg option, but also some additional statistics at the first peak and in between.

For the ESS$\nu$SB the increased dependence on the CP violating phase of the probability is well worth the loss of precious neutrino events at the second maximum. Indeed, it could provide unprecedented discovery potential to leptonic CP-violation or the most precise measurement of the corresponding phase after discovery, which could be instrumental in tackling the flavour puzzle. Moreover, as pointed out in Ref.~\cite{Coloma:2011pg} and as we will elaborate in later sections, this choice also makes the physics reach much more resilient against unexpected sources of systematic errors, since the signal, while small, has a leading dependence on the unknown parameters. Conversely, statistics will be the bottleneck of the ESS$\nu$SB physics reach and thus longer periods of data taking would greatly increase its capabilities.

On the other hand, other potential oscillation searches, different from the CP violation search, will be negatively impacted by the choice of the second oscillation maximum baseline. In particular the sensitivity to the octant of $\theta_{23}$ is severely reduced by this choice. Indeed, this measurement mainly relies on the ``atmospheric'' term of the oscillation probability, which is leading at the first maximum instead, together with $\theta_{13}$ information from reactor measurements and $\Delta m^2_{31}$ and $\sin^2 2\theta_{23}$ from $\nu_\mu$ disappearance. Similarly the $\nu_\mu$ disappearance data and hence the precise determination of $\Delta m^2_{31}$ and $\sin^2 2\theta_{23}$ are negatively affected by the choice of the second oscillation maximum. The lack of knowledge on the octant of $\theta_{23}$ can lead to ``octant degeneracies''~\cite{Fogli:1996pv} that in turn somewhat limit the CP discovery potential of the ESS$\nu$SB~\cite{Ghosh:2019sfi}. The sensitivity to the mass ordering is also limited at the ESS$\nu$SB given the small matter effects from the low energy and short baseline. However, since these matter effects are small, the resulting ``sign degeneracies''~\cite{Minakata:2001qm} do not compromise the sensitivity to $\delta$ of the facility~\cite{Baussan:2013zcy,Ghosh:2019sfi}.

A very effective and convenient way of increasing both the octant and mass ordering sensitivity of a neutrino Super Beam experiment is to combine the signal from the neutrino beam with the huge atmospheric neutrino sample that can be collected at such a detector~\cite{Huber:2005ep,Campagne:2006yx}. In the case of the ESS$\nu$SB this combination is particularly synergistic. Indeed, the atmospheric neutrino sample can provide not only significantly increased sensitivity to the octant and the mass ordering to solve parametric degeneracies, but also improved precision to $\Delta m^2_{31}$ and $\sin^2 2\theta_{23}$ which is otherwise one of the main drawbacks of the setup.  

In this work we will combine the observation of the ESS$\nu$SB flux tuned for the second maximum of the $\nu_e$ appearance probability with the complementary atmospheric neutrino data, more strongly dominated by the first maximum and $\nu_\mu$ disappearance, and characterized by stronger matter effects. We will explore how the physics reach of the facility improves when beam data is considered together with the atmospheric neutrino sample and then review the optimization of the ESS$\nu$SB facility using both data sets. 
Finally, we will discuss which sources of systematic errors among the ones considered impact the final sensitivity more significantly.

This paper is organized as follows. In Section~\ref{sec:theory} we discuss the peculiarities of the neutrino oscillation probability and the appearance of parametric degeneracies when observing at the second oscillation maximum. In Section~\ref{sec:setup} we describe the experimental setup considered and the details of the numerical simulations performed. Section~\ref{sec:results} describes the results of the simulations and in Section~\ref{sec:conclusions} we present our conclusions and summarize our work.

\section{Measurements at the second oscillation peak}
\label{sec:theory}

\begin{figure}[t]
\centering
\hspace*{-1cm}
\includegraphics[width=13.5cm]{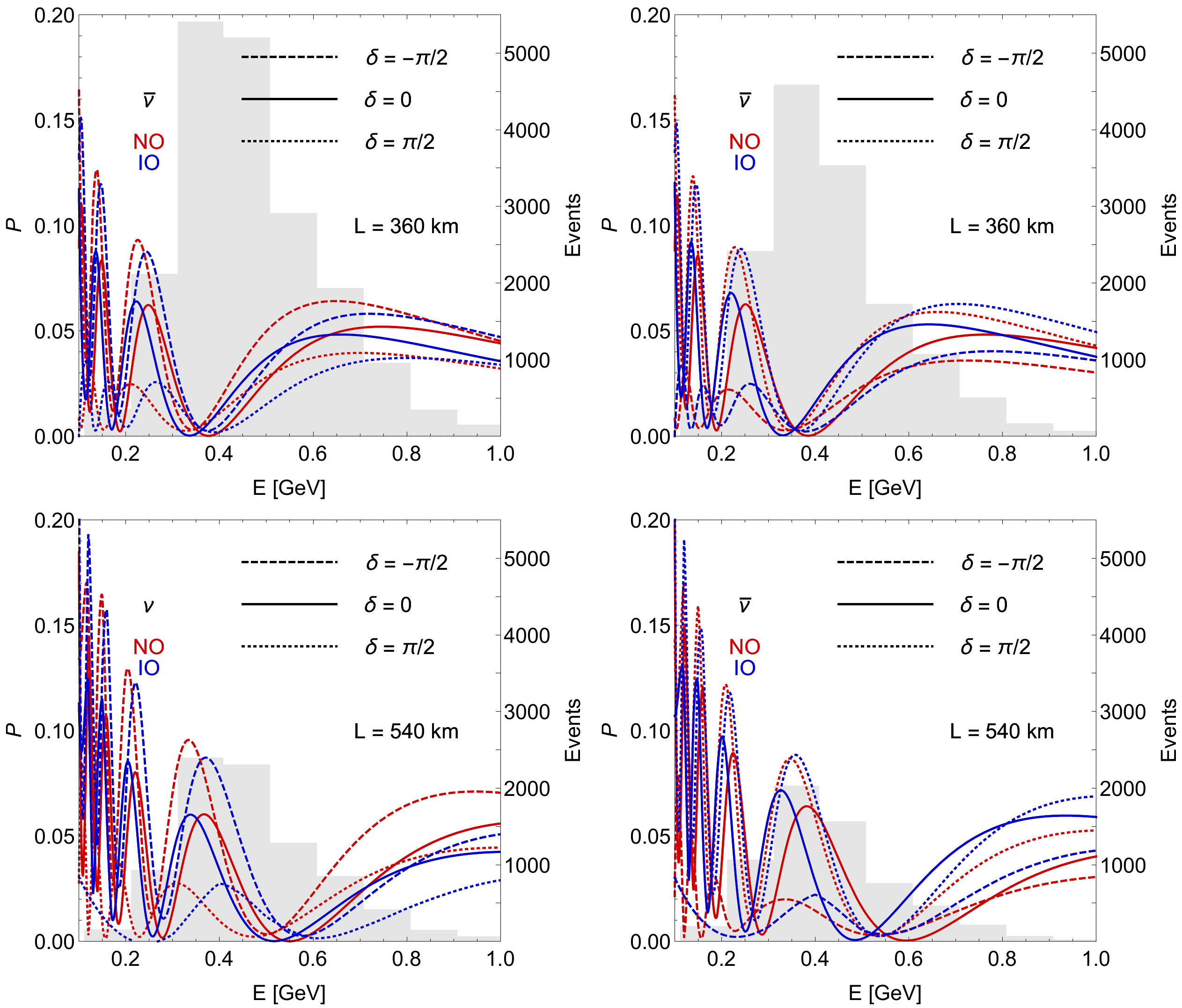}
\caption{Oscillation probabilities for the Zinkgruvan (upper panels) and Garpenberg (lower panels) baselines as a function of the energy for neutrinos (left panels) and antineutrinos (right panels). The red (blue) lines are for normal (inverted) ordering and three different values of $\delta = -\pi/2$, $0$ and $\pi/2$ are represented by the dashed, solid and dotted lines respectively. The grey histograms show the number of events that would be obtained in each energy bin for a 2/8 time splitting between neutrino/antineutrino mode if the oscillation probability was $1$. Thus, they serve as a guide of what energies of the oscillation probability would be well-sampled by the ESS$\nu$SB setup.}
\label{Fig:probs}
\end{figure}

The determination of the oscillation parameters at beam experiments is, in general, hindered by the appearance of degenerate solutions,
cf. e.g., Refs.~\cite{BurguetCastell:2001ez,Barger:2001yr,Minakata:2013hgk,Coloma:2014kca,Ghosh:2015ena}. These degeneracies have been extensively studied for the experimental setups of T2HK~\cite{Coloma:2012ji,C.:2014ika,Ghosh:2014rna,Abe:2014oxa,Ghosh:2017ged,Abe:2018uyc} and DUNE~\cite{Coloma:2012ji,Adams:2013qkq,Barger:2013rha,Ghosh:2013pfa,Agarwalla:2013vyc,Barger:2014dfa,Bora:2014zwa,Acciarri:2015uup,Nath:2015kjg,DeRomeri:2016qwo,Ghosh:2017ged,Abi:2018dnh,deGouvea:2019ozk,Ghoshal:2019pab,Meloni:2018xnk} (and also their combination~\cite{Fukasawa:2016yue,Ballett:2016daj}).
As stated in Section~\ref{sec:introduction}, the $L/E$ range which the ESS$\nu$SB focuses on is different from those of other forthcoming experiments,\footnote{
The MOMENT proposal~\cite{Cao:2014bea,Blennow:2015cmn,Bakhti:2016prn,Tang:2019wsv} with $L=150$ km can access to the oscillation probability with similar $L/E$ to the ESS$\nu$SB.
The T2HKK proposal~\cite{Ishitsuka:2005qi,Hagiwara:2005pe,Hagiwara:2006vn,Kajita:2006bt,Hagiwara:2006nn,Hagiwara:2009bb,Hagiwara:2011kw,Hagiwara:2012mg,Hagiwara:2016qtb,Abe:2016ero,Raut:2017dbh}, in which the first and the second oscillation maxima are measured with two detectors located at different sites, would also cover the similar $L/E$ range to the ESS$\nu$SB.}
Therefore, here we will discuss the peculiarities of ESS$\nu$SB and the differences from other experiments in the determination of the oscillation parameters before presenting our numerical results. The $\nu_e$ appearance oscillation probability in matter is given by~\cite{Cervera:2000kp} (see also~\cite{Freund:1999gy, Akhmedov:2004ny, Minakata:2015gra}):
\begin{equation}
    \begin{split}
        P(\barparenb{\nu}_{\mu}\rightarrow&\barparenb{\nu}_e)  = s_{23}^2\sin^2{2\theta_{13}}\left(\frac{\Delta_{31}}{\tilde{B}_{\mp}}\right)^2\sin^2{\left(\frac{\tilde{B}_{\mp}L}{2}\right)}+c_{23}^2\sin^2{2\theta_{12}}\left(\frac{\Delta_{21}}{A}\right)^2\sin^2{\left(\frac{A L}{2}\right)}\\
        & + \tilde{J}\frac{\Delta_{21}}{A}\frac{\Delta_{31}}{\tilde{B}_{\mp}}\sin{\left(\frac{A L}{2}\right)}\sin{\left(\frac{\tilde{B}_{\mp}L}{2}\right)}\left[\cos{\delta} \cos{\left(\frac{\Delta_{31}L}{2}\right)}\mp\sin{\delta}\sin{\left(\frac{\Delta_{31}L}{2}\right)}\right],
    \end{split}
    \label{Eq:Probability}
\end{equation}
where $\Delta_{i j} \equiv \Delta m^2_{i j}/ 2E$, $\tilde{J}=c_{13}\sin{2\theta_{12}}\sin{2\theta_{23}}\sin{2\theta_{13}}$, $A = \sqrt{2} G_F n_e$ is the matter potential with $n_e$ the electron density and $G_F$ the Fermi constant, and $\tilde{B}_{\mp}\equiv |A\mp \Delta_{13}|$. In this expression the only dependence in the CP violating phase $\delta$ appears in the last term, which is the interference between the ``atmospheric'' oscillation in the first term and the ``solar'' in the second. Since $\sin 2 \theta_{13} \sim 0.3$ while $\Delta_{12} L \sim 0.05$ at the first oscillation peak, the ``atmospheric'' term tends to dominate the oscillation probability and the interesting CP interference is only subleading. Conversely, at the second oscillation maximum $\Delta_{12} L \sim 0.1$ so that the dependence on $\delta$ of the oscillation probability is much higher which allows to improve the sensitivity to this parameter~\cite{Coloma:2011pg}. This can be seen in Fig.~\ref{Fig:probs} where the change in the probability upon changing the values of $\delta$ is much more significant at the second peak maximum compared to the first.

In Eq.~(\ref{Eq:Probability}) the leading dependence on the mass ordering comes from the ``atmospheric'' term, as it goes as the inverse of the square of $\tilde{B}_{\mp}$. For $E \sim |\Delta m_{31}^2|/(2A) $ there will be a resonance which will produce an enhancement in neutrinos against antineutrinos or viceversa depending on the mass ordering. For a typical average matter density of $3.0~\text{g}/\text{cm}^3$ one finds that the approximate energy for this resonance to happen is $E \sim \mathcal{O}(\text{GeV})$. Given that the peak of the flux for ESS$\nu$SB happens at $E\sim \mathcal{O}(100)~\text{MeV}$ (see Fig.~\ref{Fig:probs}), the importance of the matter effects and hence of the sensitivity to the mass ordering for this facility is not expected to be significant.

\begin{figure}[t]
\centering
\includegraphics[width=\linewidth]{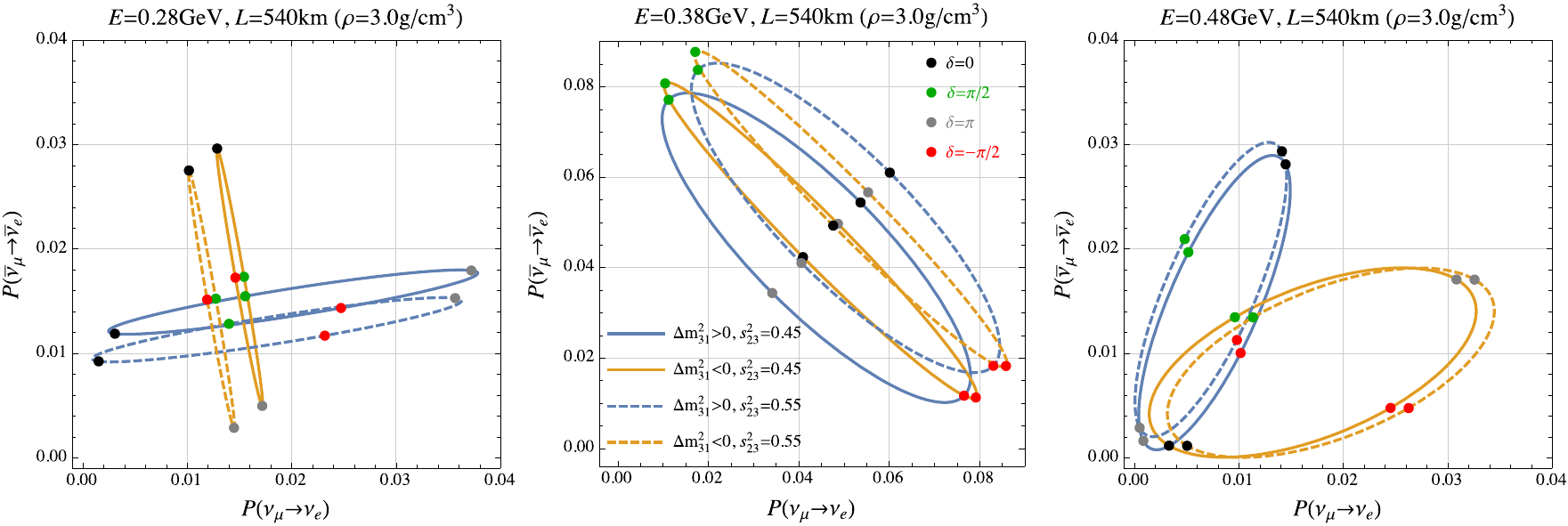}
\caption{Bi-probability plots for the ESS-Garpenberg setup $L=540$ km. Three plots for three different neutrino energies: $E=\{280, 380, 480\}$ MeV from left to right.
The four ellipses in each plot for the different choices of ($s_{23}^{2} \equiv \sin^{2}\theta_{23}$, sign[$\Delta m_{31}^{2}$]):
blue solid for ($0.45,+$), orange solid for ($0.45,-$), blue dashed for ($0.55,+$), and orange dashed for ($0.55,-$).
The energies $E=380$ MeV and $E=480$ MeV correspond to the vicinity of the second oscillation maximum and the first oscillation minimum.}
\label{Fig:biP-540}
\end{figure}

The bi-probability plots~\cite{Minakata:2001qm} shown in Fig.~\ref{Fig:biP-540} help to illustrate the degeneracy problem at the ESS$\nu$SB experiment.
Here all oscillation parameters other than $\delta$, the octant of $\theta_{23}$, and
the sign of $\Delta m_{31}^{2}$ are fixed at the current best fit
values~\cite{Esteban:2018azc}, and the matter density along the neutrino
baseline is assumed to be constant with an average density of 3.0 g/cm$^{3}$.
The baseline length $L$ and the neutrino energies $E$ are set to $L=540$ km (ESS-Garpenberg) and $E=\{280, 380, 480\}$ MeV.
The ellipses show the variation of the appearance probabilities for the
neutrino and antineutrino channels from changes in $\delta$.
The four ellipses in each plot correspond to the different choices of the octant of $\theta_{23}$ and the mass ordering.
When the ellipses overlap sharing the same region in the $P(\nu_{\mu} \rightarrow \nu_{e})$-$P(\bar{\nu}_{\mu} \rightarrow \bar{\nu}_{e})$ plane, the same oscillation probabilities can be obtained by changing $\delta$, the octant of $\theta_{23}$ and/or the mass ordering, implying the existence of degenerate solutions.

Let us first focus on the middle plot with $E=380$ MeV where the
oscillation probabilities are close to the second maximum, $|\Delta m_{31}^{2}|L/(4E) \sim 3\pi/2$.
The centres of the ellipses are located on the CP conserving line $P(\nu_{\mu} \rightarrow \nu_{e}) = P(\bar{\nu}_{\mu} \rightarrow \bar{\nu}_{e})$, which reflects the fact that the matter effect, which could induce an explicit difference between the neutrino and antineutrino oscillation probabilities unrelated to the intrinsic CP violation from $\delta$, is irrelevant for this energy and baseline.
The major axes of the ellipses extend widely along the diagonal line orthogonal to the CP conserving line. This means that the CP violating term proportional to $\sin\delta$ in Eq.(\ref{Eq:Probability}) is very relevant in the oscillation probability for this energy and baseline, leading to the improved CP sensitivity at the second oscillation peak.

\begin{figure}[t]
\centering
\includegraphics[width=\linewidth]{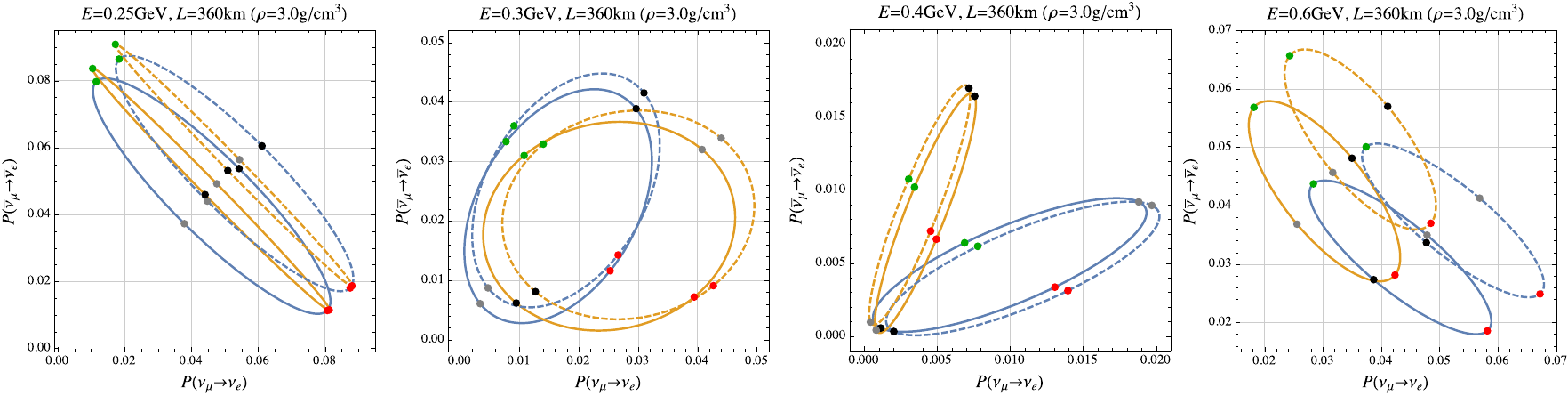}
\caption{Bi-probability plots for $L=360$ km (ESS-Zinkgruvan). In this
 energy range $E=250-600$ MeV, the oscillation probabilities
 experience the second maximum, the first minimum, and the first
 maximum.}
\label{Fig:biP-360}
\end{figure}

The ``fake'' CP violation effect due to the matter effect separates the two ellipses with opposite mass ordering at the first oscillation maximum, where T2HK focuses on, causing the $\delta$-sign$(\Delta m_{31}^{2})$ degeneracy in the CP violation search, cf. the right most plot in Fig.~\ref{Fig:biP-360}.
Conversely, the CP violation search at the second oscillation maximum is not noticeably affected by the matter effect~\cite{Bernabeu:2018use,Ghosh:2019sfi}.
Changing the value of $\theta_{23}$, the ellipses almost keep the same shape and move in parallel along the CP conserving line, which causes the $\delta$-$\theta_{23}$ degeneracy~\cite{Minakata:2013hgk,Coloma:2014kca}.

The vertices of the ellipses are located at $\delta=\{\pi/2, -\pi/2\}$, where the oscillation probabilities do not change much with a change of $\delta$. As a consequence, the precision in the determination of $\delta$ becomes worse close to the oscillation maxima~\cite{Coloma:2012wq}.
In other words, since the two points with $\delta$ and $\pi-\delta$ on an ellipse are close to each other around $\delta=\{\pi/2,-\pi/2\}$, it is hard to separate them~\cite{Coloma:2012wq}.
%
%
Although at the probability level from Fig.~\ref{Fig:biP-540} the expectation would be that this quasi-degeneracy effect occurs similarly at $\delta=\pi/2$ and $\delta=-\pi/2$, the numerical simulations we will report in Section~\ref{sec:results} show that the ESS$\nu$SB suffers this effect more severely at $\delta=-\pi/2$ than at $\delta=\pi/2$. This is due to the significant difference in event rates between these two points. Indeed, for $\delta=-\pi/2$, the oscillation probability for neutrinos is enhanced while the antineutrino one is suppressed. Since both the flux and the cross section are also smaller for antineutrinos, this strongly penalizes the measurement at $\delta = -\pi/2$ since the antineutrino sample is essentially lost given that the event rate at the second oscillation peak is already necessarily small.
On the other hand, at $\delta=\pi/2$, the oscillation probability for neutrinos is suppressed, but the larger cross section and flux compensate for it and prevents such a big loss of sensitivity. 

In the energy region that the ESS$\nu$SB focuses on, the oscillation phase changes rapidly. As a consequence, the shape and location of the ellipses changes very significantly even within the same energy bin.
In Fig.~\ref{Fig:biP-540},
we also show the bi-probability plots with $E=$280 and 480 MeV where the oscillation probabilities are approaching the minima, which are also well-covered by the ESS$\nu$SB flux.
The ellipses are not distributed symmetrically to the CP conserving line,
which means that, contrary to the second peak, matter effects do have some impact on the oscillation probabilities.
However, this impact is still subleading, given the rather low energy, and does not shift the energies where the extrema are located, cf. Fig.~\ref{Fig:probs}.
As a result, the two ellipses for the different mass hierarchies are not separated in the entire energy region.
%
%

The drastic shape change of the ellipses when varying the energy is largely due to the ratio of the $\sin\delta$ and the $\cos\delta$ terms in the oscillation probability, see Eq.~(\ref{Eq:Probability}).
The $\sin\delta$ term is most significant close to the oscillation peak with $|\Delta m_{31}^{2}| L/(4E) \simeq 3 \pi/2$ for $E \simeq 380$ MeV.
As the probabilities depart from the maximum, the major axes of the ellipses start following along the direction of the CP conserving line, which means that the $\cos\delta$ term increases in importance as we approach the minima with $|\Delta m_{31}^{2}| L/(4E) \simeq \pi$ (right panel of Fig.~\ref{Fig:biP-540}) or $|\Delta m_{31}^{2}| L/(4E) \simeq 2 \pi$ (left panel).
In the left and the right plots, the ellipses with different mass orderings intersect each other at points with different values of $\delta$ at different energies.
Therefore, in principle, with precise enough measurements at various energies, one could determine the value of $\delta$ and the sign of $\Delta m_{31}^{2}$ separately. However, the oscillations are too fast for the $\sim 100$~MeV resolution achievable at these energies with a water Cerenkov detector to resolve and also the event rate at the second maximum is not large enough to perform a very fine binning.
Thus, it is not possible to track the rapid oscillations in Fig.~\ref{Fig:probs}, although some mild sensitivity to the mass ordering can be achievable.

A large overlap between the two ellipses with different mass orderings and different octants at the oscillation maximum (middle panel in Fig.~\ref{Fig:biP-540}), where most of the statistics is concentrated, suggests that the mass ordering sensitivity at the beam experiment is affected by the octant degeneracy.

The ellipses for different octants barely separate in the entire energy region, which implies a rather poor sensitivity to $\theta_{23}$ in the appearance channel leading to octant degeneracies that can spoil both the determination of $\delta$ and of the mass ordering at the ESS$\nu$SB. Conversely, for  experiments focusing on the first maxium the two ellipses for different octants are more separated~\cite{Ghosh:2019sfi}, cf. the right panel in Fig.~\ref{Fig:biP-360}.
Therefore, we will explore the impact of the addition of the atmospheric neutrino data collected at the far detector of the ESS$\nu$SB to the beam data since atmospheric neutrinos can provide both sensitivity to the $\theta_{23}$ octant and the mass ordering helping to lift parametric degeneracies~\cite{Huber:2005ep,Campagne:2006yx}.
%

%
The mass ordering sensitivity from an observation of atmospheric neutrinos comes from the oscillation signals driven by $\Delta m_{31}^{2}$ and the matter effect (first term in Eq.~(\ref{Eq:Probability})) and therefore, it does not depend on the value of $\delta$.
On the other hand, the sensitivity is better for $\theta_{23}$ in the second octant than the first octant, since the term is proportional to $\sin^{2} \theta_{23}$~\cite{Akhmedov:2012ah}.
%

If the shorter baseline $L=360$ km (ESS-Zinkgruvan) is instead considered, the neutrino flux at the high energy tail up to $E\sim600$ MeV covers the first oscillation maximum.
This situation corresponds to the bi-probability ellipses presented in the right panel of Fig.~\ref{Fig:biP-360}, which show the same shape and position characteristic of other experiments located at the first oscillation maximum such as T2HK.
The matter effect is not significant enough to completely separate the two mass orderings.
In the relevant energy range (200-600 MeV), the oscillation probabilities go from the first maximum (right panel) to the first minimum (middle panels) and to the second maximum (left panel).
The leftmost panel with $E=250$ MeV, where the second oscillation peak would be located, looks very similar to that with $E=380$ MeV in the case of $L=540$ km.
The ellipses for the different mass orderings are separated more clearly in the case of $L=360$ km than $L=540$ km in a large energy region, which leads to a slightly better sensitivity to the mass ordering even though the baseline is shorter. 
From the information at the first oscillation maximum, the ESS$\nu$SB with $L=360$ km also has better sensitivity to $\theta_{23}$ than the $L=540$ km option, so that it is expected that the longer baseline option will benefit more from the addition of the atmospheric neutrino data, which helps to determine $\theta_{23}$ and its octant. 

\section{Simulation and experimental details}
\label{sec:setup}

The simulation of the ESS$\nu$SB data has been performed with the GLoBES software~\cite{Huber:2004ka, Huber:2007ji}. We have assumed that the neutrino beam will shine on a near and a far detector to reduce the systematic uncertainties~\cite{Baussan:2013zcy}. The far detector is a 1~Mt MEMPHYS-like water Cerenkov detector~\cite{Agostino:2012fd}, while the near detector has been assumed to be identical to the far detector in terms of efficiencies and background rejection capabilities with a fiducial mass of 0.1 kt. The response of the detectors has been implemented through migration matrices, both for the signal efficiency and the background rejection from Ref.~\cite{Agostino:2012fd}.

A beam power of 5~MW with 2.5~GeV protons and an exposure of $1.7\times 10^{7}$ operating seconds per year has been assumed~\cite{Baussan:2013zcy}. The fluxes have been simulated explicitly at 1~km for the near detector~\cite{Blennow:2014fqa}, accounting for possible geometrical effects since the source cannot be considered point-like, as well as for 100~km (and consequently rescaled) for the longer baselines considered for the far detector~\cite{Baussan:2013zcy}. The event rate peaks around $\mathcal{O}(100)$ MeV energies (see Fig.\ref{Fig:probs}), so the dominant contribution to the cross section will be in the quasi-elastic regime (QE). For the cross section we use the results from the Genie~\cite{Andreopoulos:2015wxa} tune G18$\_$10a$\_$00$\_$000.

We have assumed a total running time of 10 years. Nonetheless, we will also study the dependence of the physics reach on the relative running time spent in positive and negative focusing in order to optimize it for the measurement of CP violation. Likewise, although the preferred location of the far detector for the ESS$\nu$SB is the Garpenberg mine at 540~km~\cite{Baussan:2013zcy}, different baselines, with emphasis in the alternative Zinkgruvan option at 360~km, will be studied to address the optimal choice. Finally, we will also study how the CP discovery potential depends on the total exposure.

Throughout all the simulations we adopt the same treatment of the systematic errors from Table~\ref{Tab:Systematics} as in Ref.~\cite{Coloma:2012ji}. Unless otherwise specified, we will assume the ``Optimistic'' systematics from the first ``Opt.'' column in Table~\ref{Tab:Systematics} although we will also show how the results are affected when the more conservative ones in the second column ``Cons.'' are considered instead.
All systematics have been introduced as nuisance parameters and the results presented have been obtained minimizing the $\chi^2$ over all of them. The systematic uncertainties associated to fluxes and cross sections have been assumed to be fully correlated between near and far detector and uncorrelated between neutrino and antineutrino components and different flavours. The uncertainties on the fiducial volumes of the near and far detectors were not assumed to be correlated. Additionally, to account for the uncertainty in the cross section between the near and far detector, arising from the different flavour composition of the beam (mainly $\nu_{\mu}$ in the near site and $\nu_e$ for the signal in the far detector), a completely uncorrelated systematic is included for their ratio (last row of Table~\ref{Tab:Systematics}). Therefore, the $\chi^2$ will be given by 
\begin{equation}
    \chi^2=\text{min}_{n_{s_i}}\left(\hat{\chi}^2_{FD}[n_{s_C}]+\hat{\chi}^2_{ND}[n_{s_C},n_{s_U}] + \frac{n_{s_C}^2}{\sigma_{n_{s_C}}^2}+\frac{n_{s_U}^2}{\sigma_{n_{s_U}}^2}\right),
\end{equation}
where $\hat{\chi}^2_{FD}$ ($\hat{\chi}^2_{ND}$) corresponds to the far (near) detector and $n_{s_C}$ ($n_{s_U}$) are the correlated (uncorrelated) systematic uncertainties.

We have added to the resulting $\chi^2$ a gaussian prior with the central values and $1\sigma$ errors from Ref.~\cite{Esteban:2018azc} for ``solar'' and ``reactor'' parameters. For the ``atmospheric'' parameters we set a prior on $\sin^2{2 \theta_{23}}$ and $|\Delta m_{31}^2|$ given that the octant for $\theta_{23}$ and the mass ordering are still unknown. Since the determination of these two parameters comes primarily from atmospherics, when adding this sample to the beam data no prior has been added on $\theta_{23}$ and $\Delta m_{31}^2$.

\begin{table}
\centering
\begin{tabular} {|| c  c  c ||}
\hline
Systematics & Opt. & Cons. \\
\hline
\hline
Fiducial volume ND & 0.2\% & 0.5\% \\
Fiducial volume FD & 1\% & 2.5\% \\
Flux error $\nu$ & 5\% & 7.5\% \\
Flux error $\bar{\nu}$ & 10\% & 15\% \\
Neutral current background & 5\% & 7.5\% \\
Cross section $\times$ eff. QE & 10\% & 15\% \\
Ratio $\nu_e/\nu_{\mu}$ QE & 3.5\% & 11\% \\
\hline
\end{tabular}
\caption{Systematic uncertainties for a super beam as described in Ref.~\cite{Coloma:2012ji} for two different scenarios, the ``Optimistic'' one and the ``Conservative'' scenario where systematics are larger.}
\label{Tab:Systematics}
\end{table}

The simulation of the atmospheric neutrino sample in MEMPHYS is the one used in the analysis from Ref.~\cite{Campagne:2006yx} where the neutrino fluxes at Gran Sasso from Honda calculations~\cite{Honda:2004yz} were used. This is a conservative estimate as fluxes become larger at higher geomagnetic latitudes such as Garpenberg or Zinkgruvan. In the simulation the events are separated between fully and partially contained events in the detector and stopping from through-going muon events. The neutral current contamination in each bin was included assuming the same ratio as Super-Kamiokande between neutral-current and unoscillated charged-current events~\cite{Ashie:2005ik}. For further details on the atmospheric sample see~\cite{Campagne:2006yx}.

\section{Results}
\label{sec:results}

\begin{figure}
    \centering
    \includegraphics[width=7.5cm]{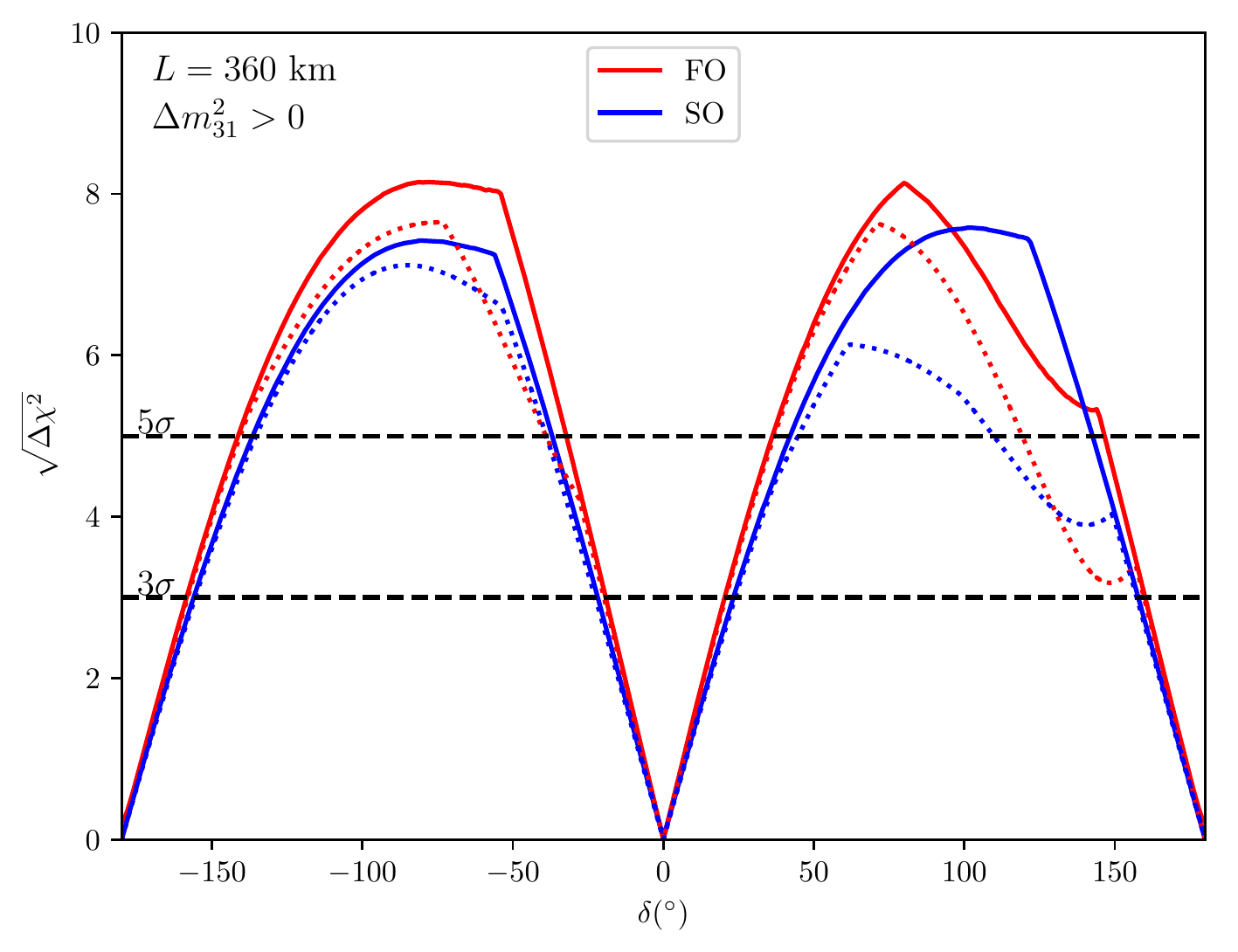}
    \includegraphics[width=7.5cm]{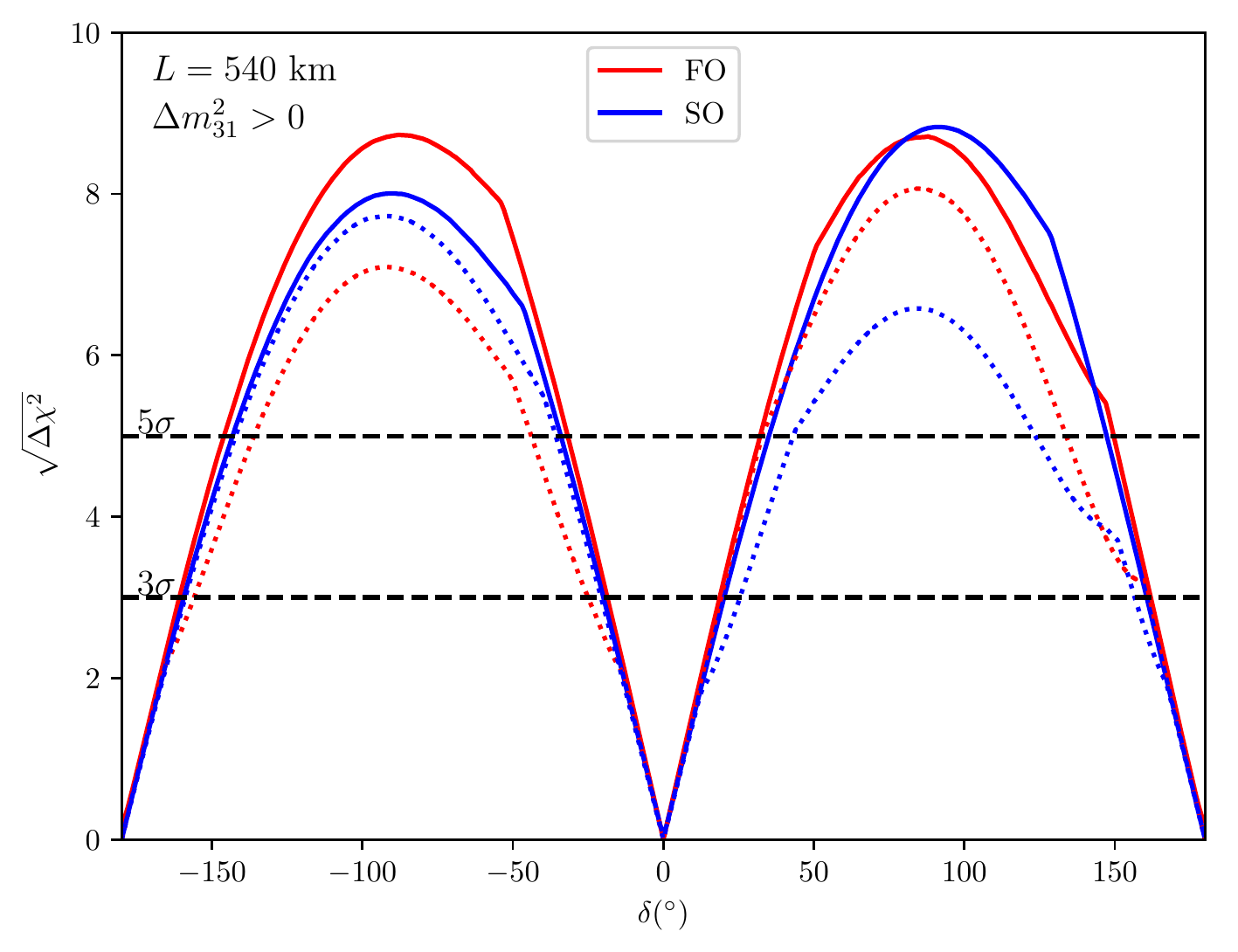}
    \includegraphics[width=7.5cm]{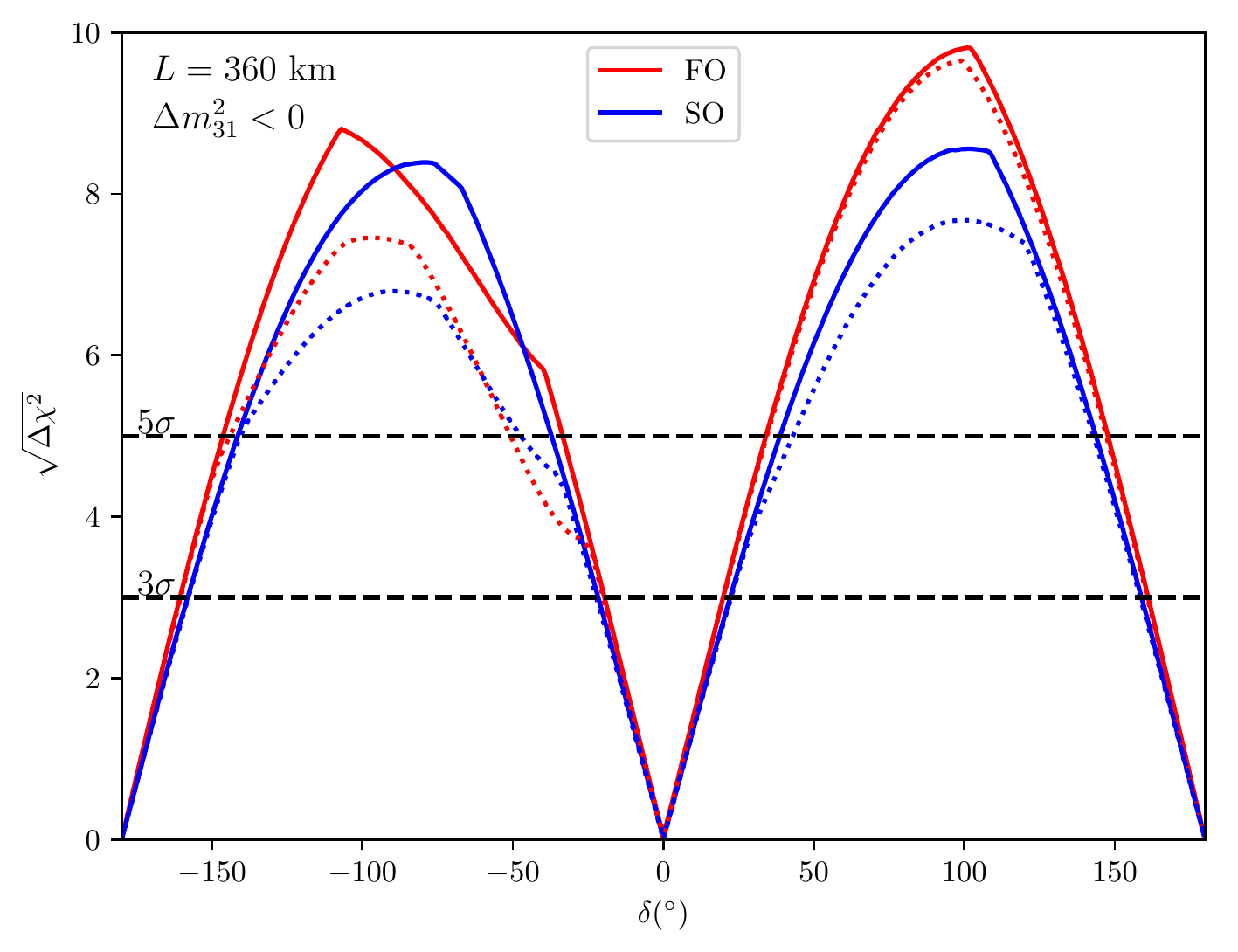}
    \includegraphics[width=7.5cm]{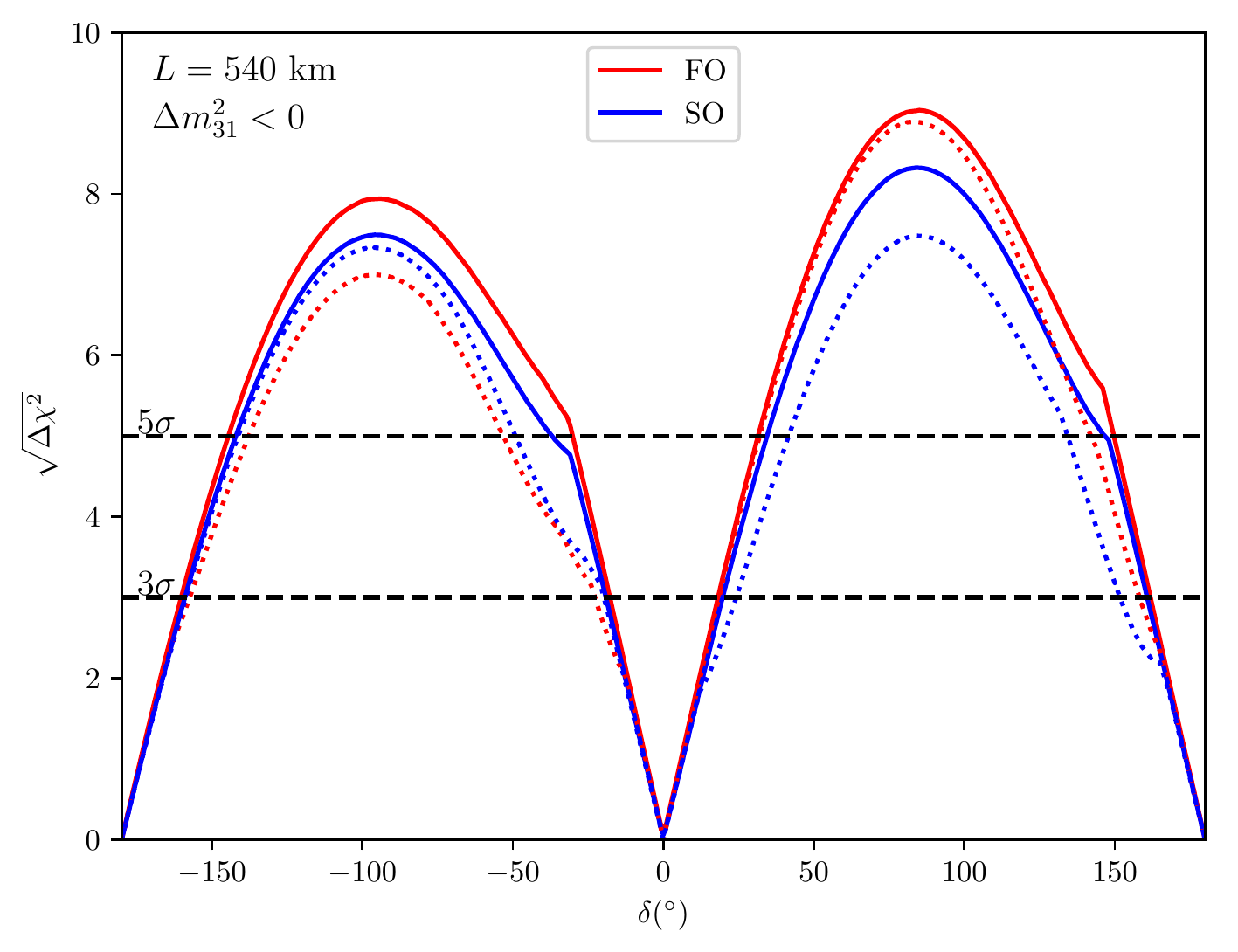}
    \caption{Significance with which CP conserving values of $\delta$ can be excluded for the Zinkgruvan 360~km (left panels) and Garpenberg 540~km (right panels) options. The upper (lower) plots are for normal (inverted) mass ordering while the red (blue) curves correspond to $\theta_{23}$ in the first (second) octant. The dashed lines correspond to the beam data only, while the continuous lines correspond to the results studying events from the beam and from atmospheric neutrinos. The running time splitting has been assumed to be $t_{\nu}$=$t_{\bar{\nu}}=5$ years.}
    \label{Fig:CP_atmvsbeam}
\end{figure}

In Fig.~\ref{Fig:CP_atmvsbeam} we show the impact on the CP discovery potential of the  ESS$\nu$SB before (dashed lines) and after (solid lines) the inclusion of the atmospheric sample for the Zinkgruvan (360~km) and Garpenberg (540~km) options in the left and right panels, respectively. The plots represent the $\sqrt{\Delta \chi^2}$ with which CP conserving values of $\delta = 0$ or $\pi$ can be disfavoured as a function of the true value of $\delta$. We take the minimum of $\Delta \chi^2$ between $\delta=0$ and $\pi$. The $\sqrt{\Delta \chi^2}$ can be interpreted as the significance for exclusion of CP-conserving values (and hence evidence for CP violation) as long as the assumptions behind Wilks' theorem hold~\cite{Wilks:1938dza}. Deviations from these assumptions can be sizable for presently running experiments, but are expected to be smaller for next generation facilities~\cite{Blennow:2014sja}.  

Even though the sensitivity of the atmospheric neutrino dataset to $\delta$ is almost negligible, the improvement of the ESS$\nu$SB physics reach upon its inclusion is quite remarkable. The improvement is generally larger for the longer 540~km baseline than for the Zinkgruvan 360~km option. This is in line with the expectations discussed in Section~\ref{sec:theory} of the atmospheric sample being more complementary to the beam information at the longer baseline. Indeed, at the second oscillation maximum the $\nu_\mu$ disappearance oscillation is not sampled as efficiently as at the first peak and this deteriorates the determination of the atmospheric oscillation parameters $\theta_{23}$ and $\Delta m^2_{31}$, which play an important role in the measurement of $\delta$. Conversely, the 360~km baseline has higher statistics and some events also cover the first oscillation maximum such that the atmospheric oscillation information is less complementary and the gain upon its inclusion is less noticeable. From these results we can conclude that the ESS$\nu$SB setup combined with the atmospheric neutrino sample would be able to rule out CP-conserving values of $\delta$ for $\sim 60 \%$  ($\sim 55 \%$) of the possible values of $\delta$ at the $5 \sigma$ level regardless of the octant and the mass ordering when observing at the 540~km (360~km) baseline. 

Figure~\ref{Fig:CP_atmvsbeam} also shows that the gain in CP discovery potential is much more pronounced in some particular regions of the parameter space, especially for $\delta < 0$ and $\theta_{23}$ in the first octant or $\delta > 0$ and the second octant. In these examples the dotted curves for beam only often show a kink that reduces the slope and the values of $\delta$ for which CP-violation could be discovered with high significance. Conversely, the corresponding solid curves with atmospheric data either do not display the kink or develop it at higher significance so that the resulting CP-discovery potential is much larger. These kinks occur due to the presence of an unresolved octant degeneracy at a CP-conserving value of $\delta$ that prevents drawing conclusions regarding CP violation. When atmospheric data is added, the sensitivity to the octant improves and these degeneracies are either lifted or only show up at much higher significance.

\begin{figure}
    \centering
    \includegraphics[width=7.5cm]{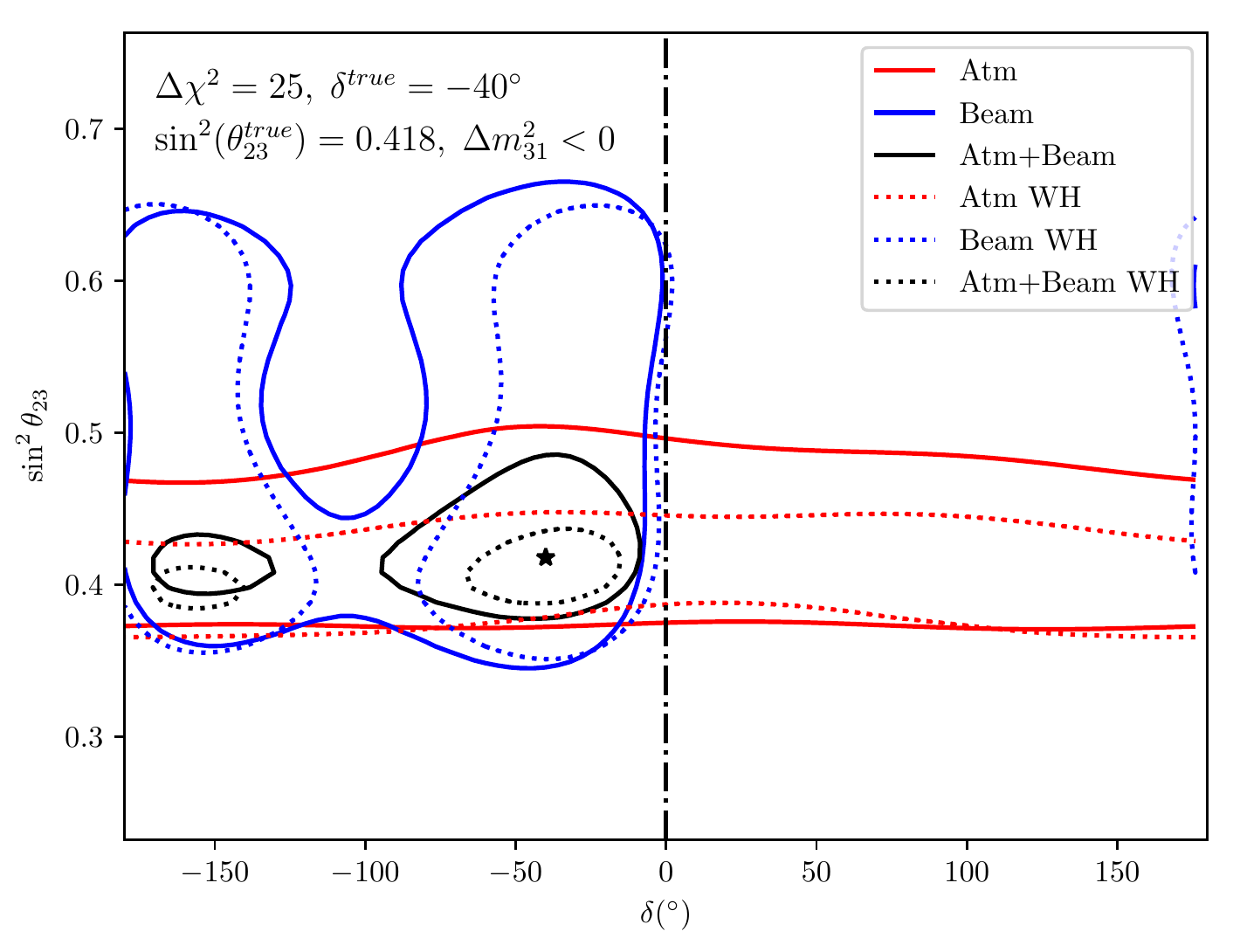}
    \includegraphics[width=7.5cm]{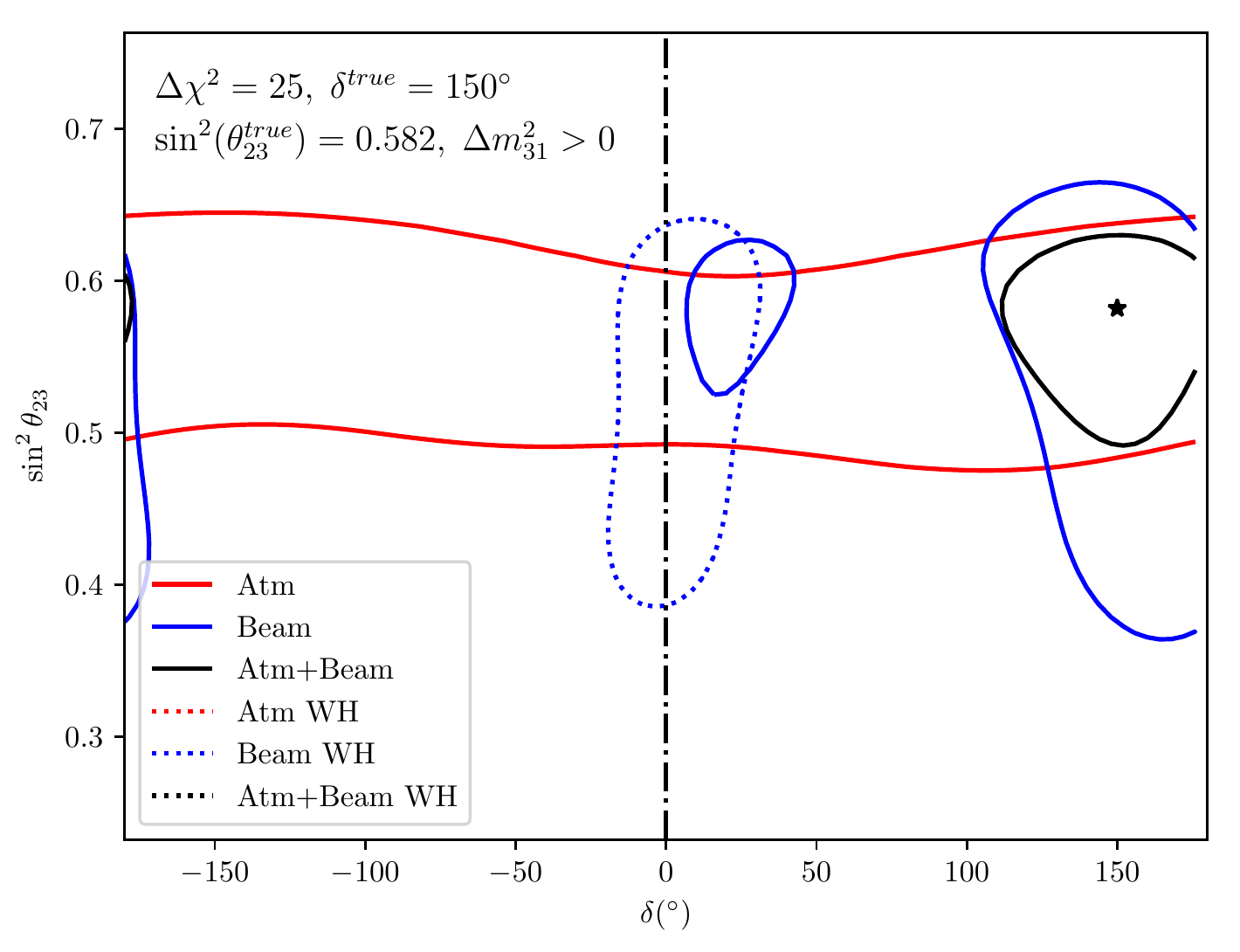}
    \caption{Allowed regions at $\Delta \chi^2 = 25$ for different assumed values of $\sin^2\theta_{23}$ and $\delta$ represented by the star for a 540~km baseline (Garpenberg location). The red curves correspond to the atmospheric dataset alone, the blue to the beam-only information and the black curves to the combination of both. Dotted regions are allowed with the wrong mass ordering. The running time splitting has been assumed to be $t_{\nu}$=$t_{\bar{\nu}}=5$ years.}
    \label{Fig:CP_sens_potato}
\end{figure}

This situation is illustrated in Fig.~\ref{Fig:CP_sens_potato}, where the allowed regions at the $\Delta \chi^2 = 25$ level are shown in the $\delta$-$\sin^2 \theta_{23}$ plane. The left (right) panels assume the true values $\delta=-40^\circ$ ($\delta=150^\circ$), $\sin^2 2 \theta_{23}=0.418$ ($\sin^2 2 \theta_{23}=0.582$) and normal ordering. As can be seen, when only the beam information is taken into account (blue curves), an octant degeneracy that spreads the allowed region towards CP conserving values appears. Conversely, the atmospheric data on their own (red curves) have no capability to determine $\delta$ at all, but can instead rule out the wrong octant of $\theta_{23}$. Thus, the combination of the two data sets (black curves) very significantly improves the CP discovery potential of the facility in these areas of parameter space. The dotted lines correspond to ``sign'' degeneracies with the opposite mass ordering to the one chosen as true value. In the right panel this degeneracy is also solved with atmospheric data while for the values of $\delta$ and $\theta_{23}$ chosen in the left panel a small sign degeneracy remains between the 4 and $5 \sigma$ level. Notice that an ``intrinsic degeneracy''~\cite{BurguetCastell:2001ez} at $\delta \simeq \pi-\delta_{true}$ also shows up at the $5 \sigma$ level when only the beam information is taken into account. As for the ``sign'' degeneracy, the atmospheric neutrino data is enough to lift it for the parameters chosen in the right panel while a small remnant is present in the left. In any case, both the ``intrinsic'' and the ``sign'' degeneracies appear at $\delta \simeq \pi-\delta_{true}$, given the comparatively small matter effects for the setup, and their allowed regions are smaller or comparable to that of the true solution so that only the ``octant''degeneracy plays a significant role in reducing the CP-discovery potential when atmospheric data is not exploited to lift it.

\begin{figure}
    \centering
    \includegraphics[width=10.5cm]{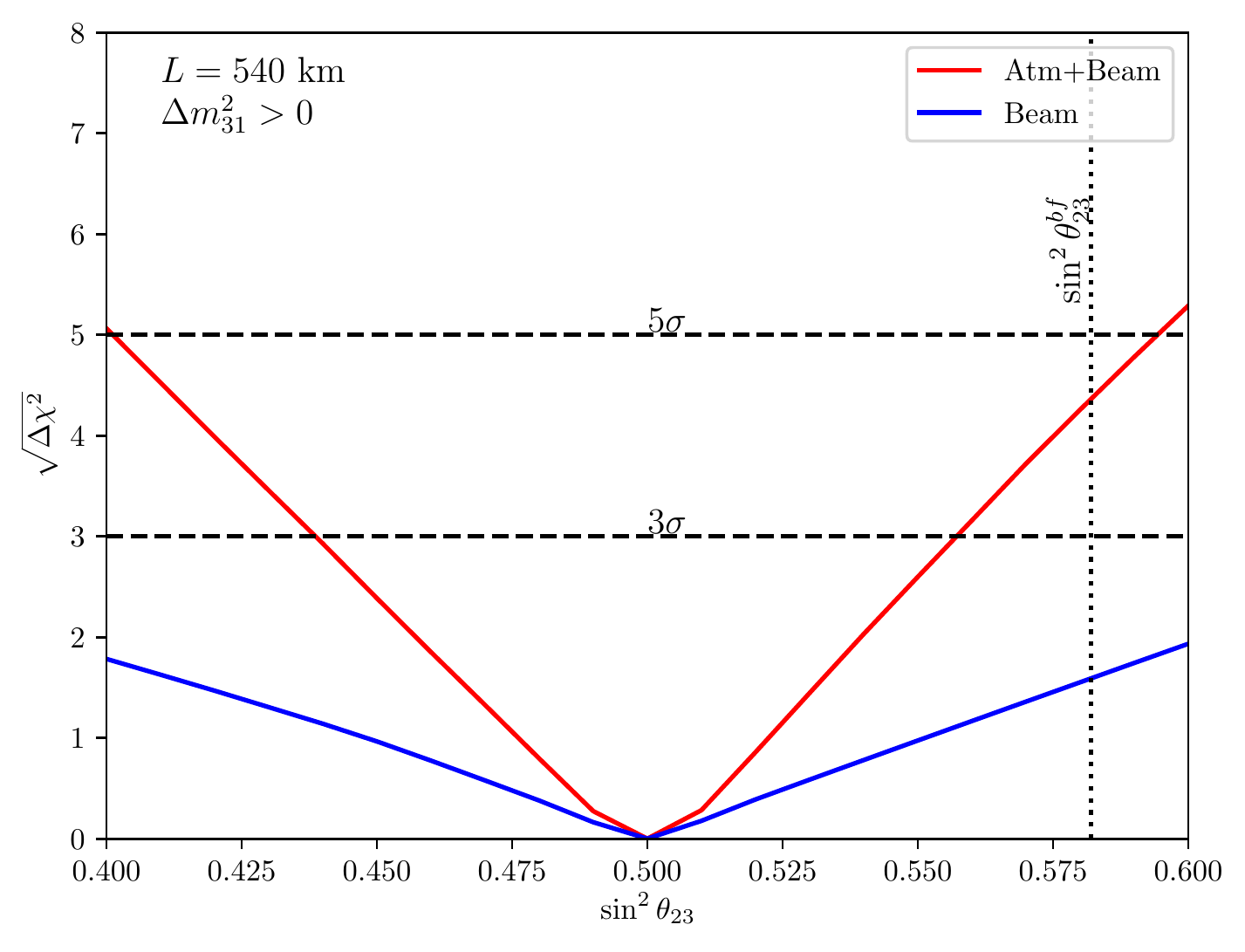}
    \caption{Significance with which the wrong octant would be disfavoured as a function of the actual value of $\theta_{23}$ with beam-only information (blue lines) and including also the atmospheric dataset (red lines) for the baseline to Garpenberg ($L=540$~km) and normal mass ordering. The running time splitting has been assumed to be $t_{\nu}$=$t_{\bar{\nu}}=5$ years. The results for the Zinkgruvan site ($L=360$~km) and for inverted ordering are very similar. The vertical line represents the present best fit for $\theta_{23}$ from~\cite{Esteban:2018azc}.}
    \label{Fig:Oct_sens}
\end{figure}

In Fig.~\ref{Fig:Oct_sens} we show how the significance with which the ESS$\nu$SB would be able to disfavour the wrong octant of $\theta_{23}$ as a function of the true value of $\theta_{23}$ (blue lines). As already anticipated in Section~\ref{sec:theory}, this capability improves dramatically upon the inclusion of the atmospheric neutrino sample (red lines) and thus the potentially dangerous ``octant'' degeneracies are lifted. The curves are almost identical for both mass orderings and for the Zinkgruvan and Garpenberg baselines.

\begin{figure}[h]
    \centering
    \includegraphics[width=7.5cm]{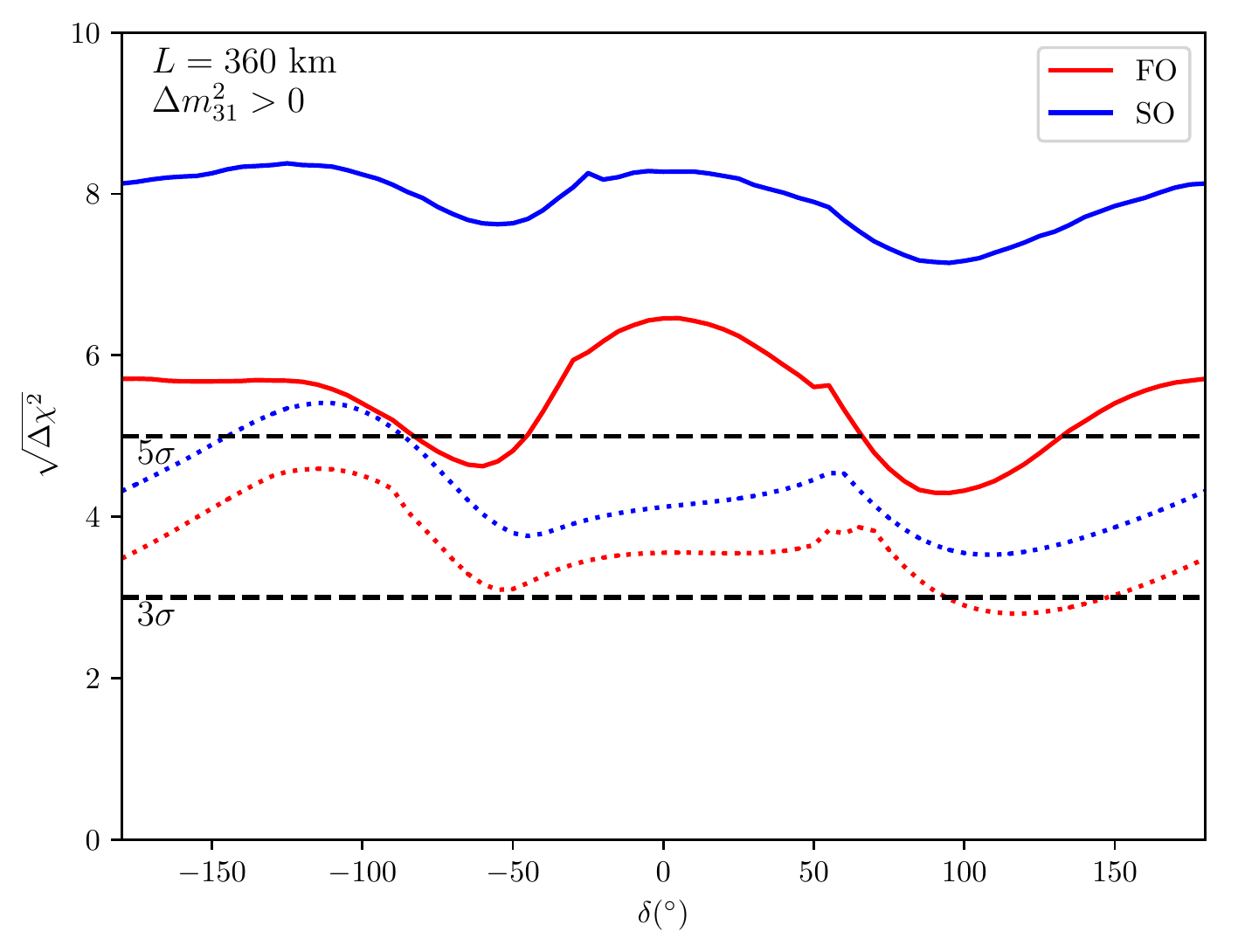}
    \includegraphics[width=7.5cm]{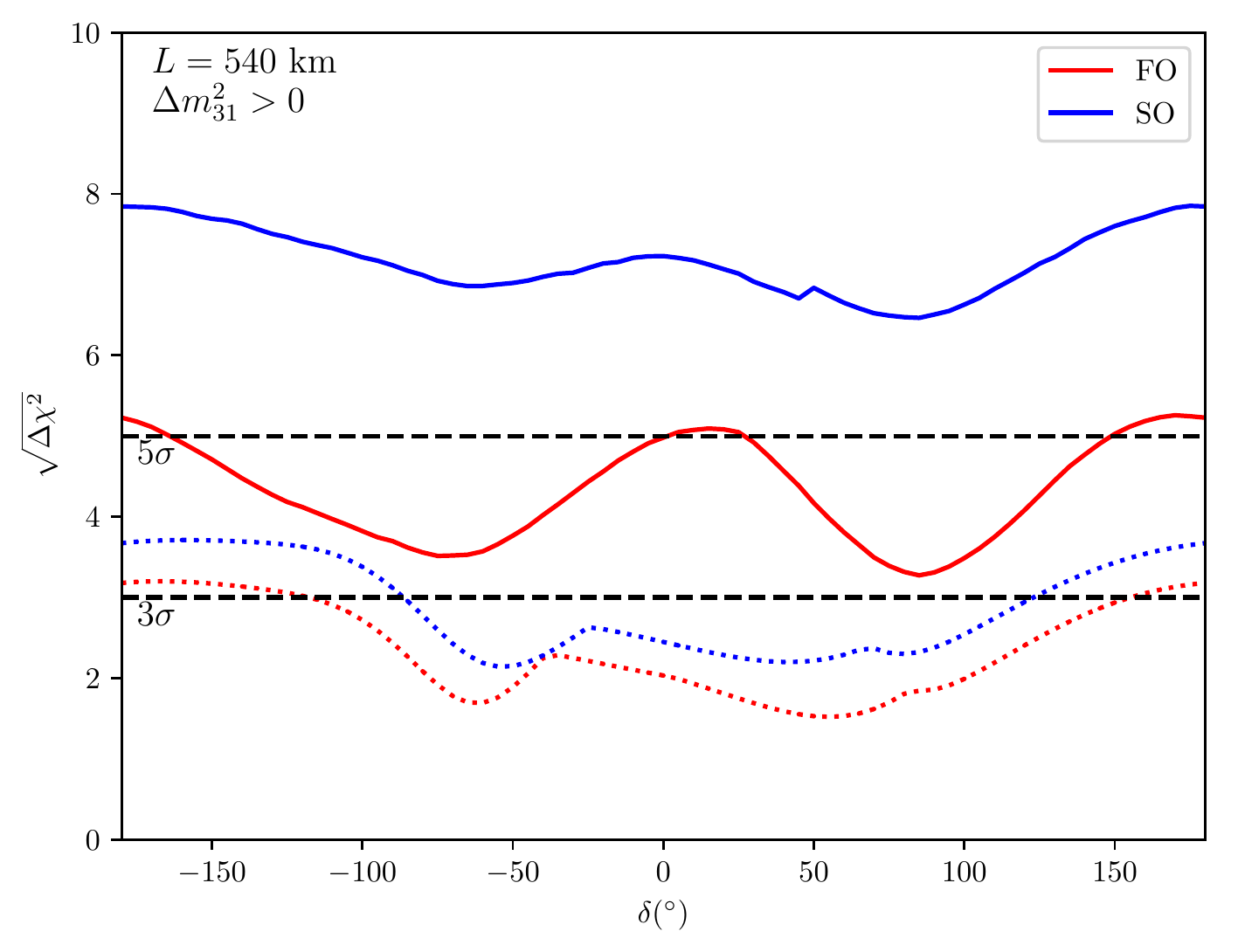}
    \includegraphics[width=7.5cm]{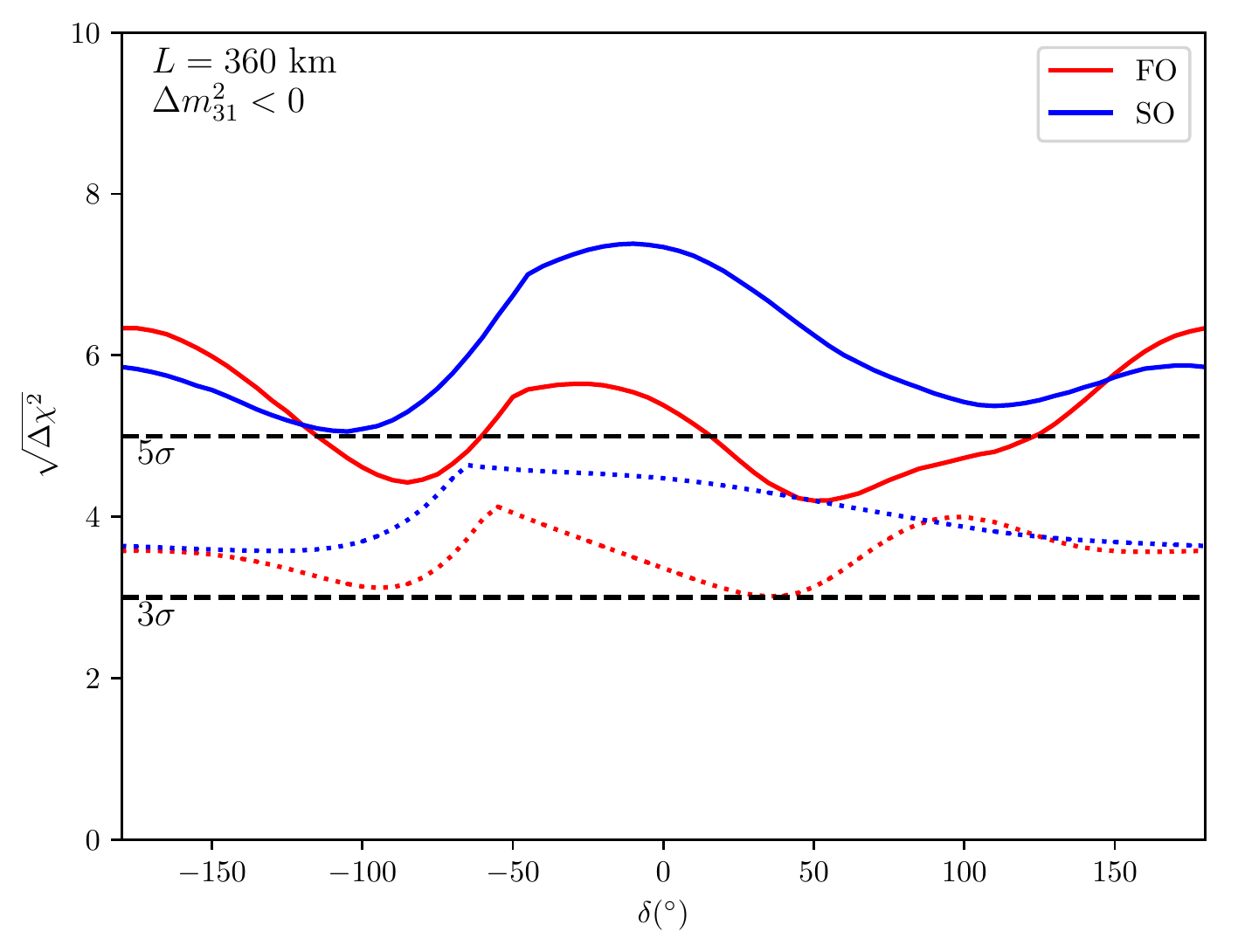}
    \includegraphics[width=7.5cm]{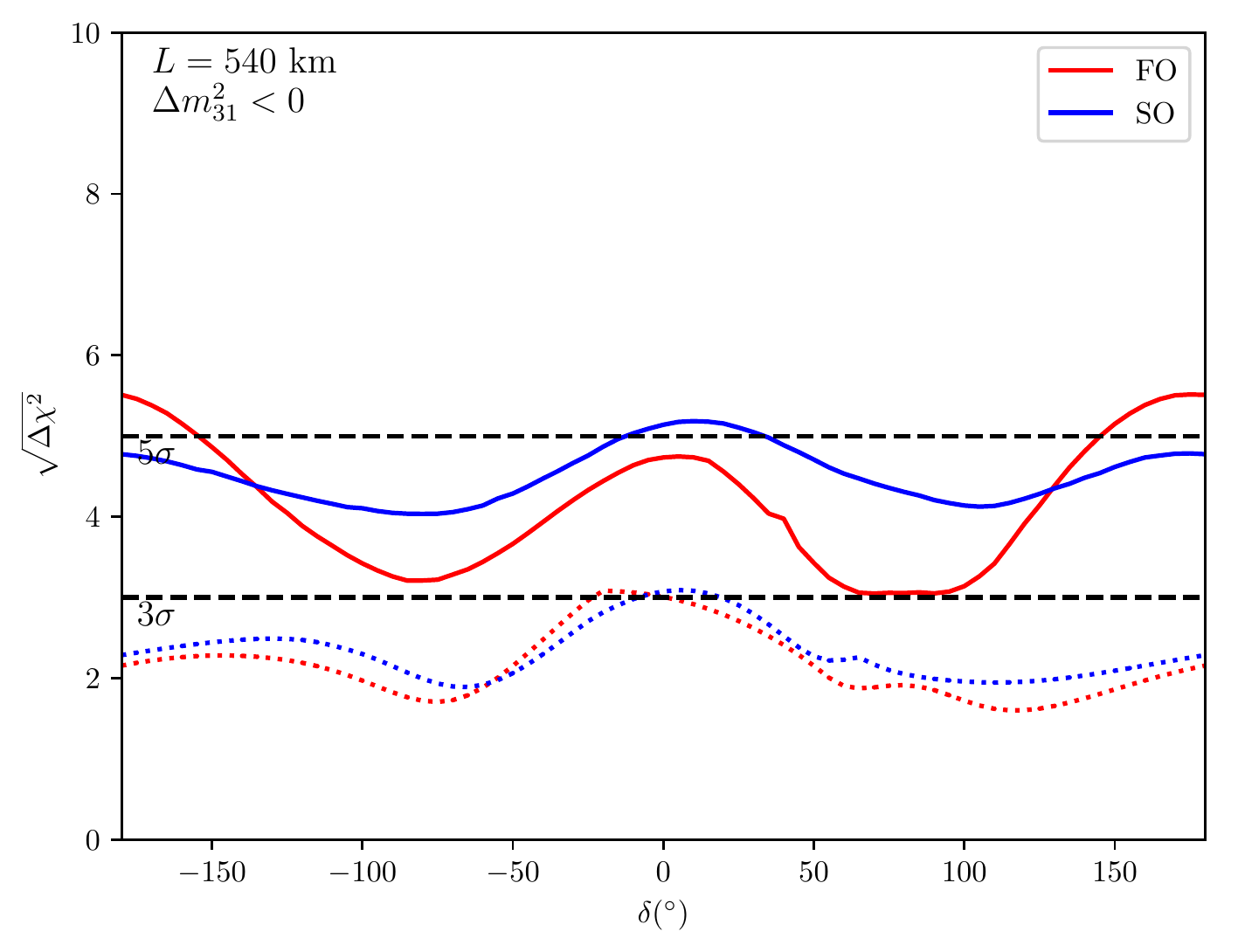}
    \caption{Significance with which the wrong mass ordering would be disfavoured for $\theta_{23}$ in the first octant (red lines) or second octant (blue lines) and the true mass ordering being normal (upper plots) or inverted (lower plots). Dashed lines correspond to the beam only data while solid lines correspond to the addition of the atmospheric sample. The left panels correspond to the baseline to Zinkgruvan while the right ones to the location of the Garpenberg mine. The running time has been assumed to be $t_{\nu}$=$t_{\bar{\nu}}=5$ years.}
    \label{Fig:Hierarchy_sens}
\end{figure}

The significance with which the ESS$\nu$SB would be able to disfavour the wrong mass ordering is shown in Fig.~\ref{Fig:Hierarchy_sens}, where dotted (solid) lines correspond to beam only data (beam and atmospheric data). The left (right) panels correspond to the 360~km (540~km) baseline and upper (lower) panels are for the scenario in which the true ordering is normal (inverted). As can be seen the ESS$\nu$SB beam data allows to disfavour the wrong mass ordering at around the $3 \sigma$ ($2 \sigma$) level  for the 360~km (540~km) baseline for any value of $\delta$ and the octant. When the atmospheric data is added, the sensitivity to the wrong ordering is boosted to the 4-5$\sigma$ level or even higher for the particular case of normal ordering and second octant of $\theta_{23}$ ($\sin^2{\theta_{23}}=0.582$ from Ref.~\cite{Esteban:2018azc}) for which the signal in atmospheric neutrinos is enhanced, as expected from Eq.(\ref{Eq:Probability}). For normal ordering (upper panels) the inclusion of the atmospheric neutrino data also change the shape of the curve, in particular a larger increase in the significance is seen around $\delta=0$ than for other values. This is due to the solution of the octant degeneracy since, as can be seen in the middle panel of Fig.~\ref{Fig:biP-540} or the first panel of Fig.~\ref{Fig:biP-360}, for $\delta=0$ and normal ordering the ellipse with opposite octant and ordering has a significant overlap.

\begin{figure}[ht]
    \centering
    \includegraphics[width=7.5cm]{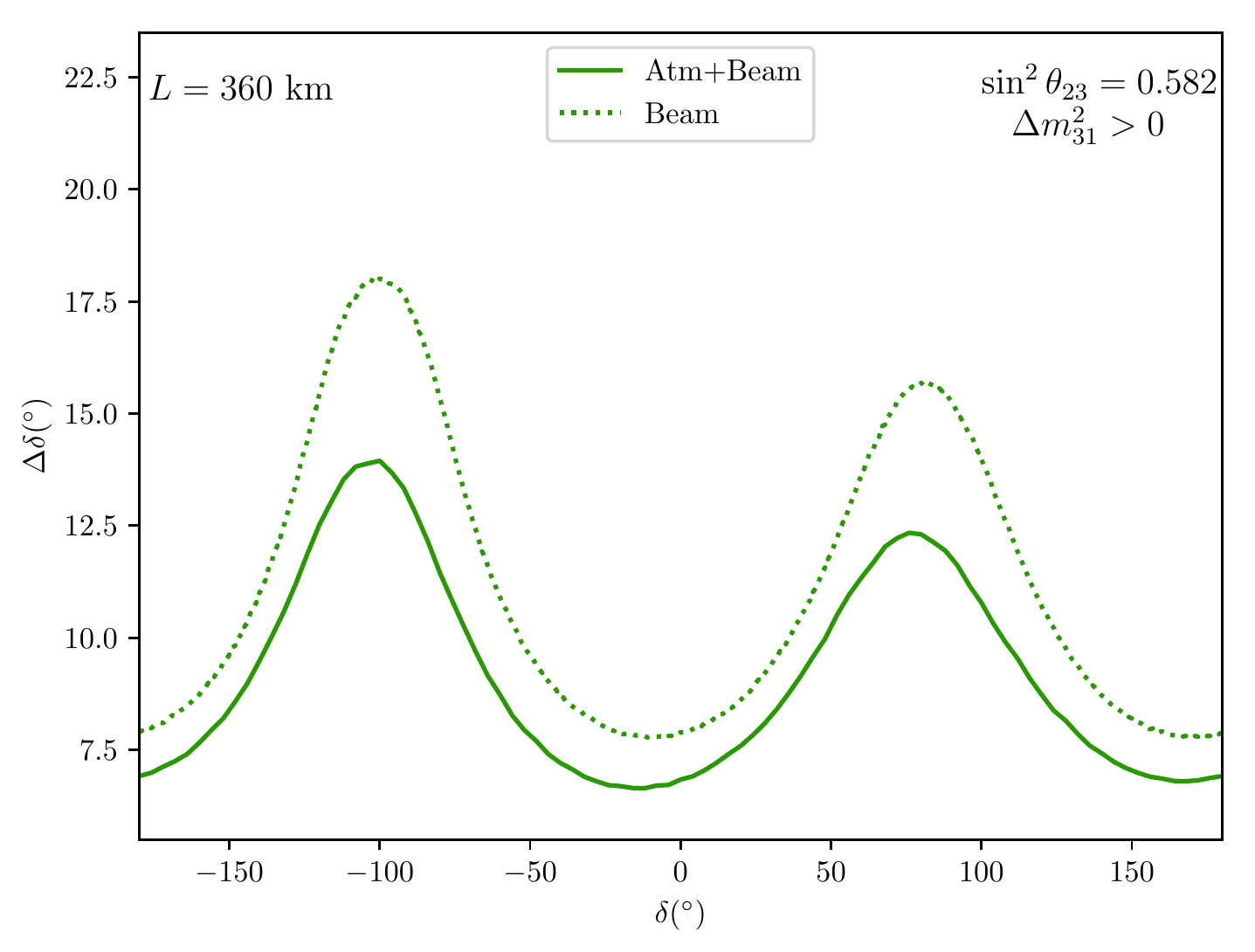}    
    \includegraphics[width=7.5cm]{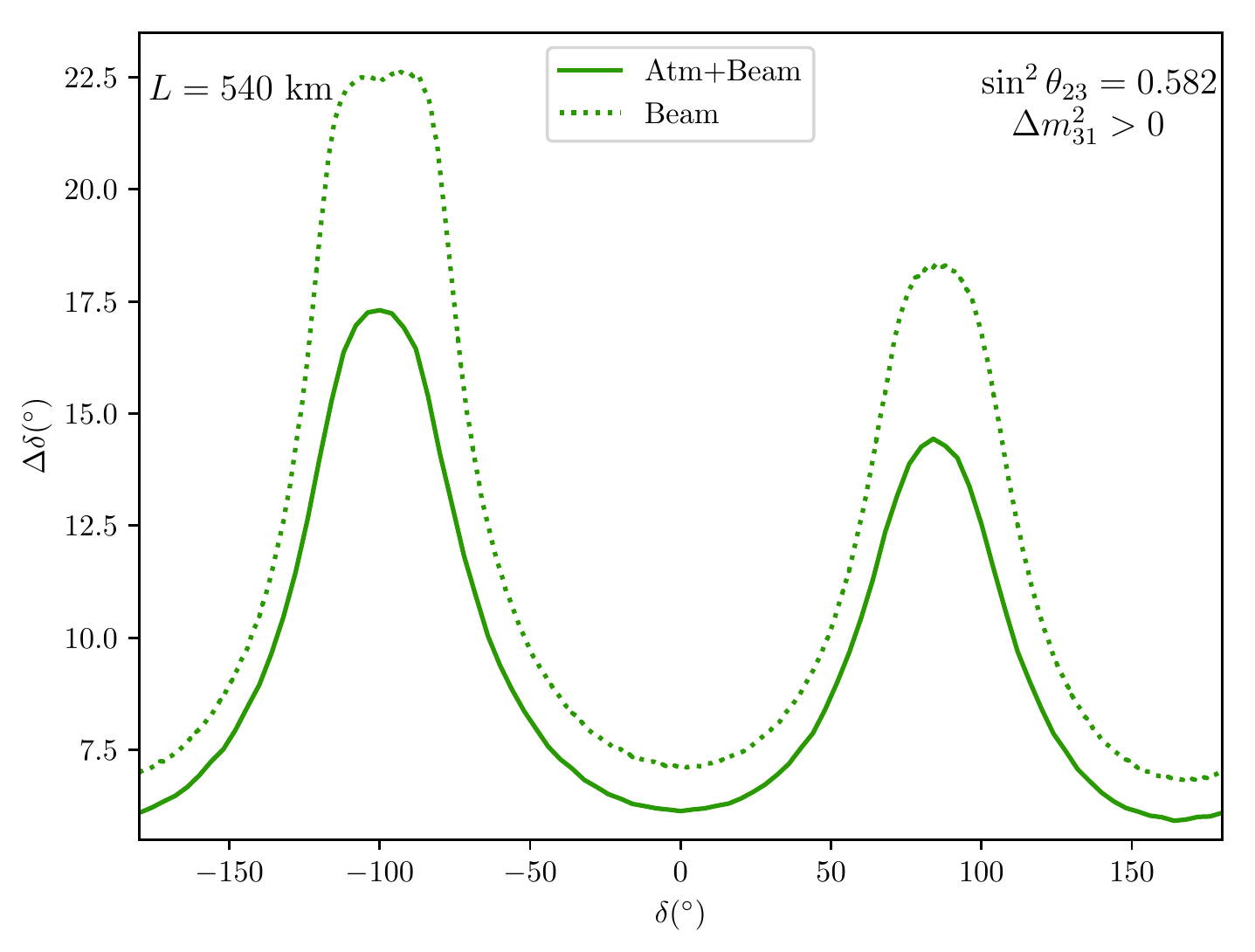}
    \includegraphics[width=7.5cm]{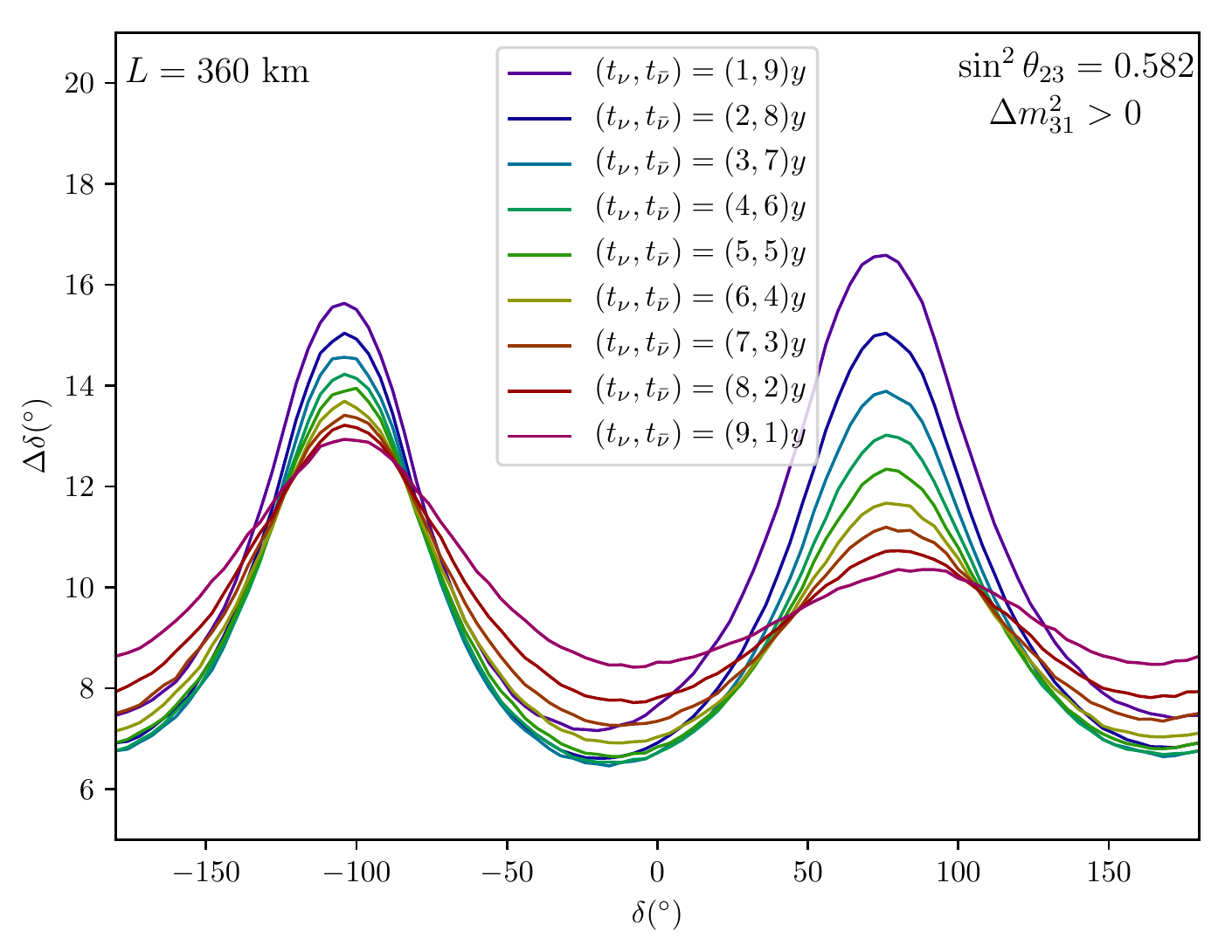}
    \includegraphics[width=7.5cm]{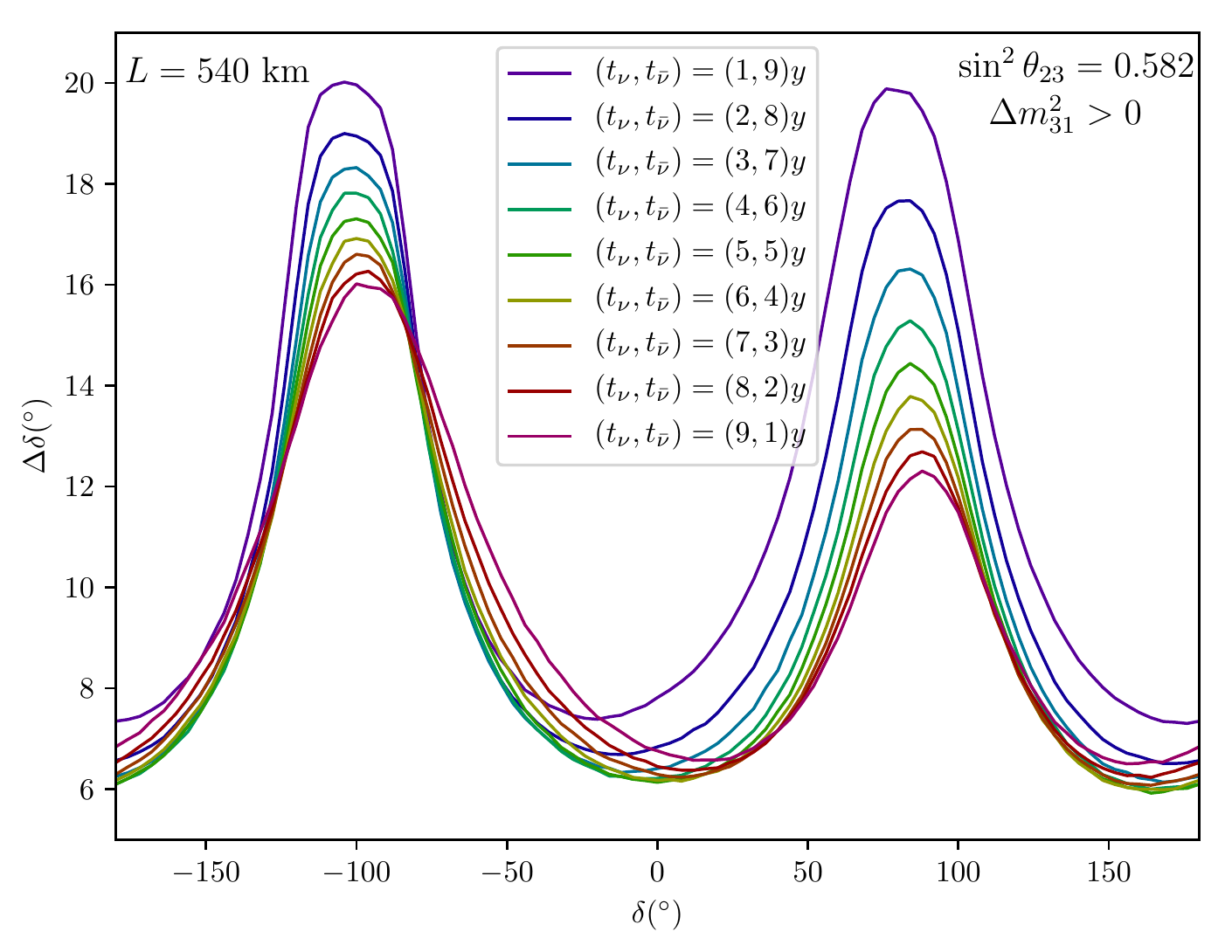}
    \caption{Precision (spread of the $1 \sigma$ allowed region) on the determination of $\delta$ for the baseline to Zinkgruvan $L=360$~km (left panels) and Garpenberg $L=540$~km (right panels) for the current best-fit parameters~\cite{Esteban:2018azc}. In the upper panels we show the comparison between the precision obtained with (solid lines) and without (dashed lines) the atmospheric sample for a running time of 5 years in each focusing. In the lower plots we show the dependence of the precision on the relative running time in each mode, where $t_{\nu}$ ($t_{\bar{\nu}}$) corresponds to the time the experiment would run in neutrino (antineutrino) mode, combining atmospheric and beam datasets.}
    \label{Fig:Precision_RunTime}
\end{figure}

In Fig.~\ref{Fig:Precision_RunTime} we analyze the precision with which the ESS$\nu$SB experiment would be able to measure the CP-violating phase $\delta$. In this figure we assumed the currently preferred option of normal ordering and second octant of $\theta_{23}$. In the upper panels we show the improvement in the $1 \sigma$ allowed region with which $\delta$ would be constrained by adding the atmospheric neutrino sample (solid lines) to the beam information alone (dotted lines). As can be seen, both for the 360~km (left panel) and 540~km baseline (right panel), the precision with which $\delta$ could be determined has a very pronounced shape. For CP violating values of $\delta$ around $\pm 90^\circ$, the $1 \sigma$ uncertainty in the measurement peaks leading to the poorest precision, while for $\delta$ around $0$ or $180^\circ$ the most precise measurements would be achieved. 

As discussed in Ref.~\cite{Coloma:2012wq}, this structure follows from the dependence of the oscillation probability on $\delta$ shown in  Eq.(\ref{Eq:Probability}). At an oscillation peak $|\Delta m_{31}^{2}| L/(4E) = (2n-1)\pi/2$ and thus mainly $\sin \delta$ is probed. Since the derivative of $\sin \delta$ vanishes at $\delta = \pm 90^\circ$, the precision with which $\delta$ can be determined is worst close to these values. In order to constrain $\delta$ around $\delta = \pm 90^\circ$, measurements away from the oscillation maxima to determine $\cos \delta$ would instead be necessary. These off-peak measurements are easier at the Zinkgruvan 360~km baseline since the statistics is higher and also the beam is not exactly centered at the maximum, while they are very challenging at Garpenberg since very few events away from the oscillation peak are expected. This explains why the reconstructed sensitivities around $\delta = \pm 90^\circ$ are much worse in the right panel compared to the left. Moreover, the double-peak structure that can be seen for  $\delta = - 90^\circ$ for 540~km corresponds to the ``intrinsic'' degeneracies depicted in Fig.~\ref{Fig:CP_sens_potato} that merge into one bigger allowed region. Since, as seen in Fig.~\ref{Fig:CP_sens_potato}, the addition of atmospheric data can lift these degeneracies, in the solid lines where this information was included the difference between the two baselines is significantly reduced.

Conversely, for $\delta = 0$ or $180^\circ$ the measurement on peak is what allows to determine $\delta$ and, since this is better covered at the longer 540~km baseline, the precision is slightly better there. This fact also translates into the better CP-discovery potential observed for the 540~km baseline in Fig.~\ref{Fig:CP_atmvsbeam}. Since the error in $\delta$ is smaller around CP-conserving values, the 540~km option could get closer to these values but still allow to claim the discovery of CP violation with high significance. 

In the lower panels of Fig.~\ref{Fig:Precision_RunTime}, the impact of changing the relative running times in positive focusing (neutrino mode) and negative focusing (antineutrino mode) is shown. Since off-peak measurements are required for $\delta = \pm 90^\circ$, statistics are crucial and easier to accumulate in neutrino mode, since fluxes and cross sections are larger, and thus the best precision would be obtained by devoting longer periods of data taking to positive focusing. Conversely, around $\delta=0$ or $180^\circ$ the complementarity between the neutrino and antineutrino samples pays off and more even splits of the running time provide better sensitivity. 

\begin{figure}
    \centering
    \includegraphics[width=7.5cm]{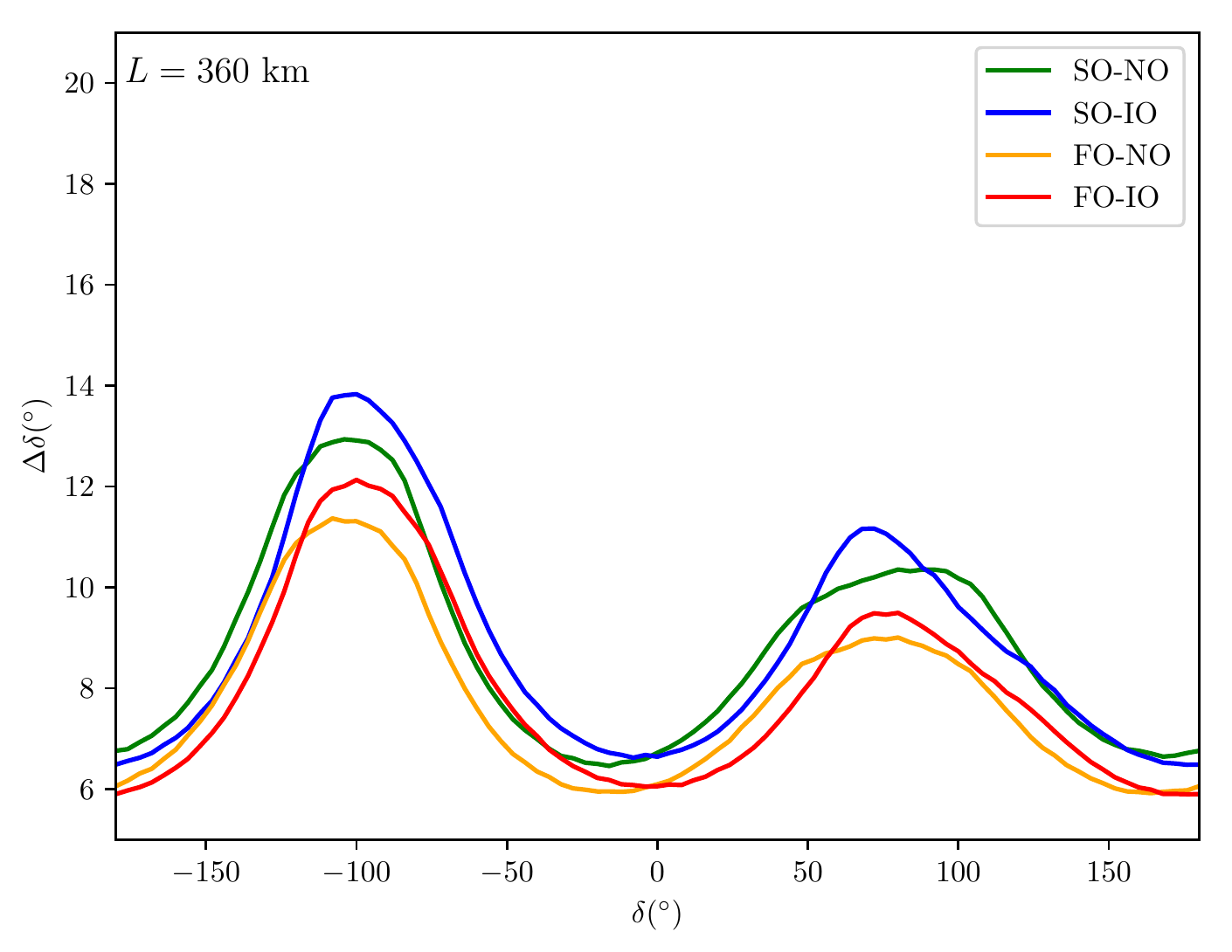}
    \includegraphics[width=7.5cm]{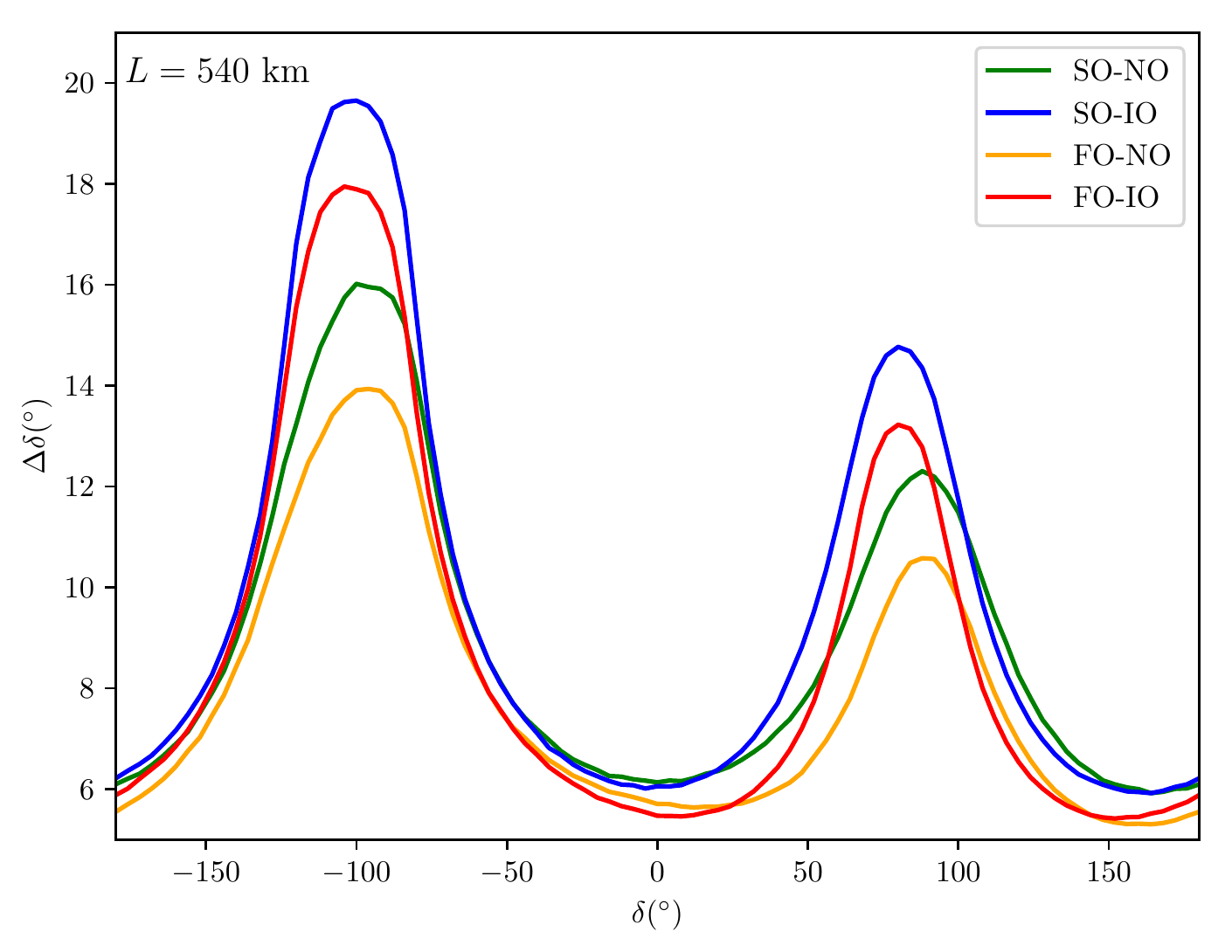}
    \caption{Precision on the measurement of $\delta$ for a total running time of 10 years when the relative running time in neutrino and antineutrino modes is optimized for each value of $\delta$. This corresponds to running similar times in neutrino and antineutrino modes around $\delta=0, 180^\circ$ and maximizing the neutrino runs around $\delta = \pm 90^\circ$.}
    \label{Fig:Precision_Best}
\end{figure}

Since the ESS$\nu$SB would be a next-generation facility, its measurement strategy can profit from the previous hints by preceding oscillation experiments and adapt the splitting between neutrino and antineutrino modes depending on what value of $\delta$ data point to. If such a strategy is followed and the best splitting between neutrino and antineutrino modes is adopted for each value of $\delta$, the precision presented in Fig.~\ref{Fig:Precision_Best} would be obtained. If the mass ordering is confirmed to be normal and $\theta_{23}$ lies in the second octant as present data prefer, the precision with which the ESS$\nu$SB facility would determine $\delta$ ranges from $16^\circ$ ($13^\circ$) for $\delta \sim -90^\circ$ to $6^\circ$ ($7^\circ$) for $\delta \sim 0$ or $\delta \sim 180^\circ$ for 540~km (360~km).

From Figs.~\ref{Fig:CP_atmvsbeam} and~\ref{Fig:Precision_Best} one can conclude that if the experiments preceding the ESS$\nu$SB do not find any evidence for CP-violation, the best option would be the 540~km baseline and a more or less even split of the neutrino and antineutrino running times. Indeed, this choice would minimize the errors with which $\delta$ would be determined around CP-conserving values and allow to increase the CP-discovery potential. On the other hand, if the previous set of experiments determine $\delta$ to be close to maximally CP-violating, then the best scenario for the ESS$\nu$SB would be the shorter 360~km baseline and increased neutrino run time to determine $\delta$ with the best precision possible. 

\begin{figure}
    \centering
    \includegraphics[width=10.5cm]{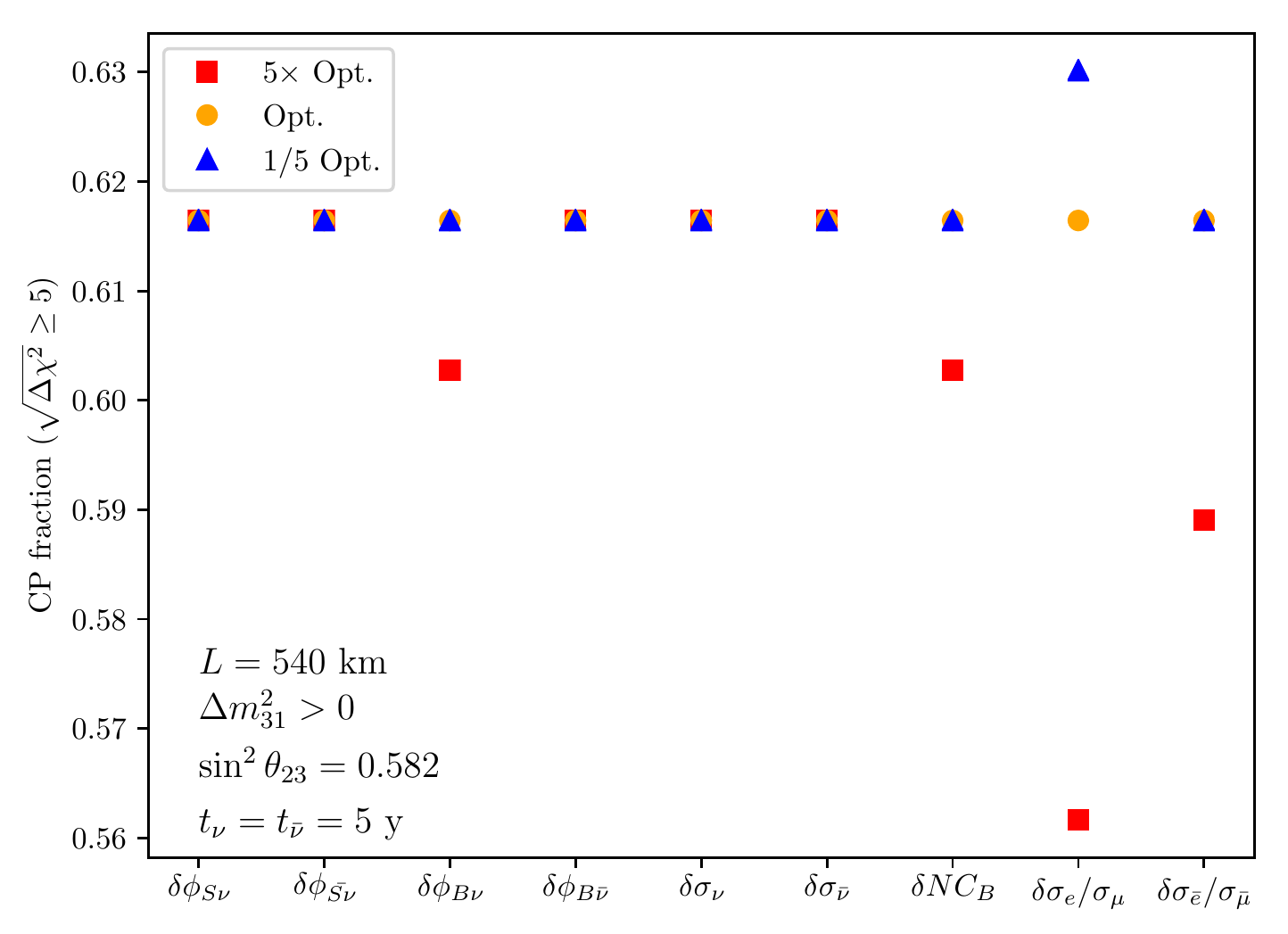}
    \caption{Impact of different sources of systematic errors on the fraction of values of $\delta$ for which a $\Delta \chi^2 > 25$ exclusion of CP conservation would be possible at the Garpenberg mine. The orange circles correspond to the CP fraction with the ``Optimistic'' systematics from Table~\ref{Tab:Systematics}, red squares correspond to assuming that particular uncertainty to be 5 times larger and blue triangles to reducing the uncertainty by a factor of 5.}
    \label{Fig:CP_frac_Sys}
\end{figure}

In Fig.~\ref{Fig:CP_frac_Sys} we show the impact of individual systematic uncertainties on the fraction of values of $\delta$ for which CP violation could be discovered ($\Delta\chi^2\geq 25$). The sources of uncertainty considered, summarized in Table~\ref{Tab:Systematics}, are the flux uncertainties for the signal ($\delta\phi_S$) and background ($\delta\phi_B$), the cross section systematic ($\delta\sigma$), the neutral current background ($\delta NC_B$), and the uncertainty on the ratio of the electron and muon flavour neutrino cross section ($\delta \sigma_e/\sigma_{\mu}$). The plot shows that the systematic uncertainties that most significantly affect the performance of the ESS$\nu$SB are the ones related to the background components of the beam, since for these the determination at the near detector is more challenging. Namely, $\delta\phi_B$, $\delta NC_B$ as well as $\delta \sigma_e/\sigma_{\mu}$ since the only $\nu_e$ present at the near detector that would allow to fix this parameter are those from the intrinsic background contamination of the beam. Among these, the strongest impact on the sensitivity is due to the cross section ratio since, not only it is difficult to constrain, but it is also most relevant to the signal at the far detector, which consists of $\nu_e$.  Indeed, reducing or increasing this particular source of systematic error has the biggest impact on the physics reach. The impact is in any event limited, since the main bottleneck to the performance when observing at the second oscillation peak is statistics. In particular, a reduction of this systematic by a factor of 5 improves the CP fraction by $\sim 2\%$ (no impact for $\bar{\nu}$) while the same factor in the opposite direction worsens the sensitivity by $\sim 9\%$ ($\sim 4\%$).

\begin{figure}
    \centering
    \includegraphics[width=7.5cm]{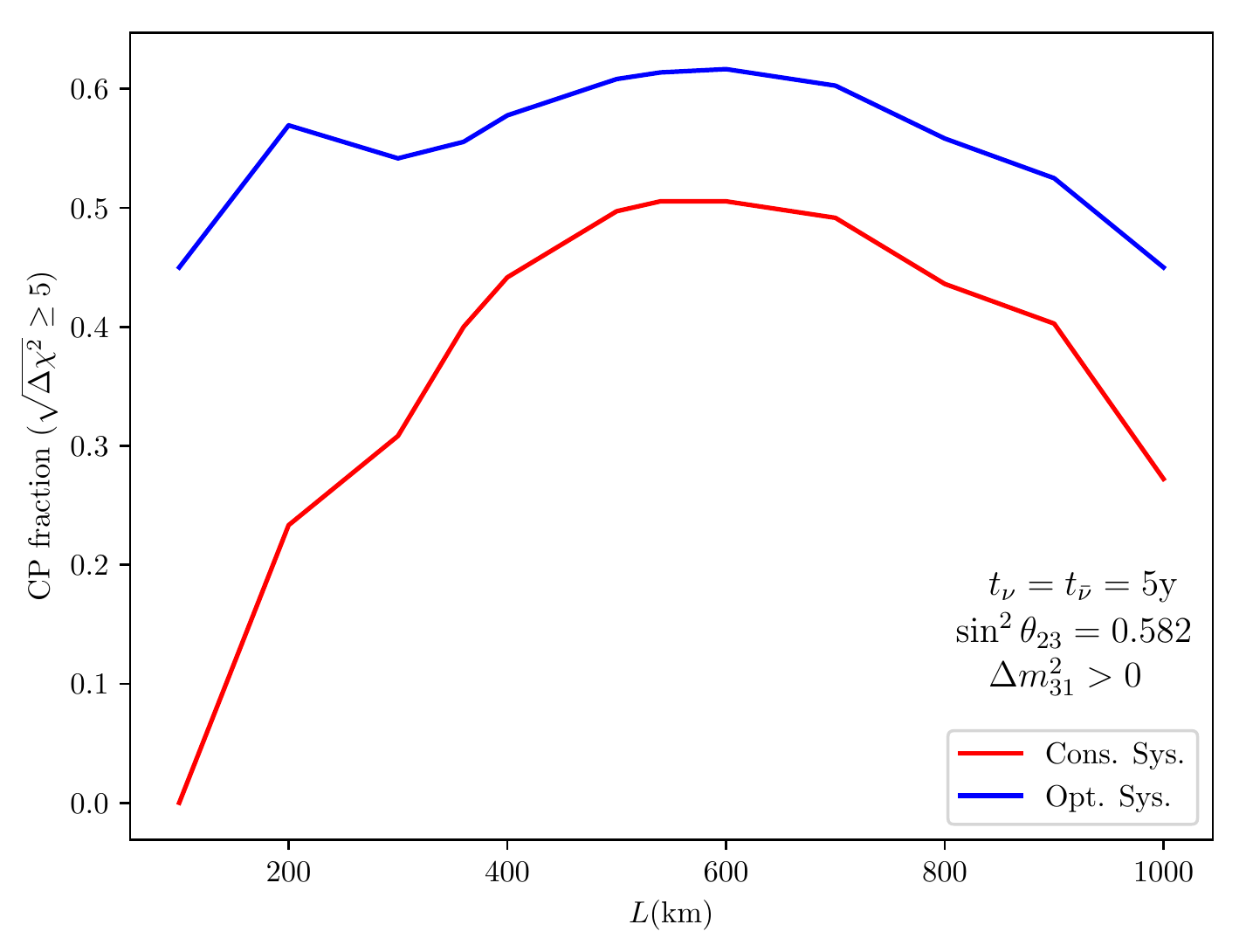}
    \includegraphics[width=7.5cm]{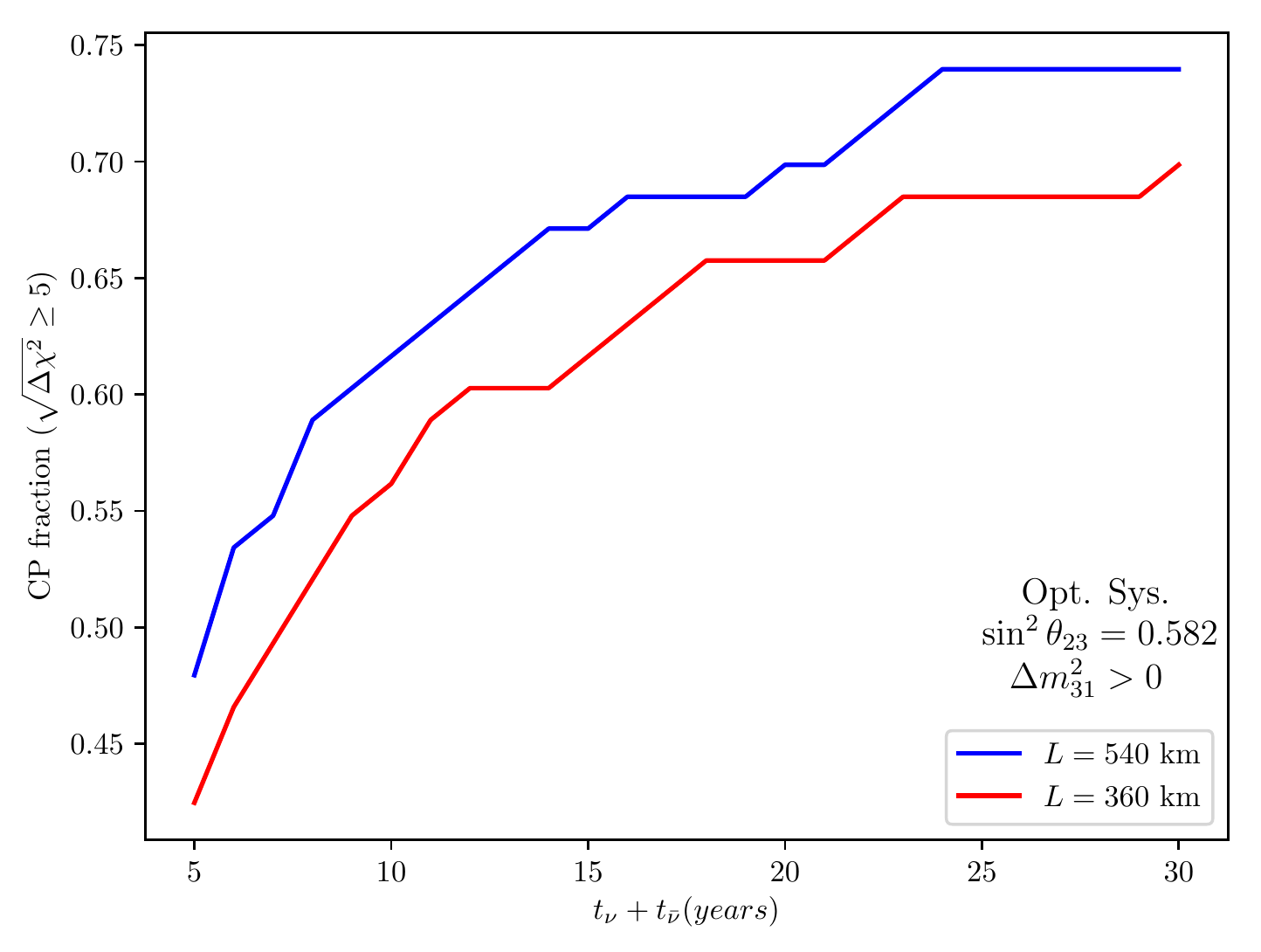}
    \caption{Fraction of values of $\delta$ for which CP violation could be discovered above $5\sigma$ for different baselines to the far detector (left panel) for the two different sets of systematics from Table~\ref{Tab:Systematics}. In the right panel we show the CP fraction for the Garpenberg ($L=540$~km) and Zinkgruvan ($L=360$~km) mines, assuming the current best fit values for the oscillation parameters and the ``Optimistic'' systematics for increasing total exposure.}
    \label{Fig:CP_fraction_Opt}
\end{figure}

The importance of these systematic errors in the physics reach is crucially dependent on the baseline of the experiment. In the left panel of Fig.~\ref{Fig:CP_fraction_Opt} we show the fraction of all the possible values of $\delta$ for which it would be possible to rule out $\delta =0$ or $\delta = 180^\circ$ with a $\Delta \chi^2 = 25$ or higher significance. The upper blue line is for the more optimistic systematics from Table~\ref{Tab:Systematics} and the lower red one for the more conservative values. As can be seen, the fraction of values of $\delta$ at which a $5 \sigma$ discovery would be possible, peaks between 400~km and 700~km in both cases. But this peak is much more pronounced when the more conservative values are assumed for the systematic uncertainties. Indeed, for larger values of the systematics, the shorter baselines are strongly penalized since the dependence of the oscillation probability is subleading around the first peak and easily hidden by the systematics. Conversely, if very small systematic errors can be achieved, then the main limiting factor would be statistics and shorter baselines would perform better. Thus, by measuring at the second oscillation maximum the ESS$\nu$SB setup becomes much more resilient to sources of systematic errors unaccounted for than when observing only at the first peak.

In the right panel of Fig.~\ref{Fig:CP_fraction_Opt} we show how the fraction of values of $\delta$ for which CP violation would be discovered at the $5 \sigma$ level by the ESS$\nu$SB beam and atmospheric data increases with the exposure. As expected from an observation at the second oscillation peak, statistics is the main factor controlling the final reach of the experiment. Indeed, for 5 years data taking the CP fraction is around $46\%$, by 10 years it increases to $62\%$ and reaches $70\%$ for 20 years of exposure. The slope only flattens significantly after 25 years.

\section{Conclusions}
\label{sec:conclusions}

In this paper we have performed an exhaustive analysis of the physics reach of the ESS$\nu$SB facility exploring its capability to determine all the presently unknown neutrino oscillation parameters such as the mass ordering and the octant of $\theta_{23}$ but with a focus on the discovery of leptonic CP violation and a precision measurement of $\delta$, which are the main declared goals of the experiment. For the first time we combined the atmospheric neutrino sample that would also be observed at the facility with the beam information and studied the complementarity between the two data sets. We studied how the physics reach of the facility could be optimized by exploring different baselines and focusing on the two candidate sites of Zinkgruvan at 360~km and Garpenberg at 540~km. We have also explored how the time split between neutrino and antineutrino modes can be exploited to improve the physics reach.

We conclude that the inclusion of the atmospheric data set can significantly increase the ESS$\nu$SB physics reach. Due to the peculiarities of observing the oscillation probability at the second oscillation maximum we find that this combination is particularly synergistic. The atmospheric neutrino sample not only significantly increases the sensitivity to the mass ordering, like for other similar facilities~\cite{Huber:2005ep,Campagne:2006yx}, but it is also very effective in improving the constraints on $\Delta m^2_{31}$ and $\theta_{23}$ and its octant. These measurements are especially challenging for the beam alone when sitting at the second maximum, given the low statistics, particularly in antineutrinos and in the $\nu_\mu$ disappearance channel. However, the determination of $\delta$ can be affected by correlations with $\theta_{23}$~\cite{Coloma:2014kca} and degeneracies with the wrong octant and thus the atmospheric information is also crucial to increase the CP discovery potential of the ESS$\nu$SB indirectly. We find this complementarity is somewhat more pronounced for the longer 540~km baseline since there the flux is more centered at the second oscillation peak and the statistics are smaller so it benefits more from the information gained from the atmospheric neutrino data.

Regarding the optimal baseline, we find the choice is rather dependent of the actual value of $\delta$. For $\delta \sim \pm 90^\circ$ a precise measurement needs events away from the oscillation maximum. In this sense the shorter 360~km baseline is better since the statistics for off-peak events are higher and this leads to a more precise measurement. Conversely, if $\delta$ is close to CP conserving values and the previous set of measurements have not been able to claim the discovery of CP-violation, the longer 540~km baseline would allow to cover a larger part of the parameter space. Indeed, after 10 years of data taking, the fraction of values of $\delta$ for which a $5 \sigma$ discovery would be possible is $56\%$ for Zinkgruvan and $62\%$ for Garpenberg.

As for the splitting of the data taking time between neutrino and antineutrino modes, the optimal strategy also depends on the value of $\delta$. This fact could be exploited since previous and present data at the time of the measurement should already show a strong preference for some part of the parameter space. Thus, the running strategy can be adapted to the situation optimizing the precision with which this measurement can be performed. In particular we find again that given the need of going beyond measurements at the peak for $\delta \sim \pm 90^\circ$, statistics is much more relevant and maximizing the time in neutrino mode translates to the best precision for these values. Conversely, close to CP-conserving values of $\delta$, the information from events on-peak is most relevant and the complementarity between neutrino and antineutrino modes pays off so that a more even split of the running time would provide the best precision.

Finally we explored the possible bottlenecks for the physics reach of the facility exploring how it is affected by varying the values of the different systematic errors considered as well as the total exposure. As expected, the choice of observing the oscillation probability at its second maximum significantly reduces the impact of the systematic errors. We find that around the first oscillation peak the fraction of values of $\delta$ for which a $5 \sigma$ discovery is possible is reduced by more than a factor 2 when considering the more conservative values of Table~\ref{Tab:Systematics}. On the other hand, at the second peak the reduction is only by a factor around $1.2$. Among the different sources of systematic uncertainties considered, the most important is the possible difference in the ratio of the electron to muon neutrino cross sections. This uncertainty is difficult to constrain from near detector information since the flux is mainly composed of $\nu_\mu$, but the far detector signal consists of $\nu_e$. Conversely, the observation at the second maximum considerably reduces the number of events and statistics play a much more relevant role. At the longer 540~km baseline, the fraction of values of $\delta$ allowing for a discovery would go from $47 \%$ to $62 \%$ and $70 \%$ for data taking periods of 5, 10, and 20 years, respectively. 

\section*{Acknowledgements}
We are extremely grateful to Michele Maltoni for providing us with the simulations of atmospheric neutrino dataset that would be collected at the MEMPHYS detector used in Ref.~\cite{Campagne:2006yx}. We are also indebted to Budimir Klicek and Marco Roda for suggestions and help with the GENIE tunes most appropriate for the ESS$\nu$SB energy range. We also want to thank WP6 of the ESS$\nu$SB design study, in particular Monojit Ghosh, for comments on our manuscript.
f
This work is supported in part by the European Union's Horizon 2020 research and innovation programme under the Marie Sklodowska-Curie grant agreements 674896-Elusives, 690575-InvisiblesPlus, and 777419-ESSnuSB, as well as by the COST Action CA15139 EuroNuNet. 
MB, EFM, and SR acknowledge support from the ``Spanish Agencia Estatal de Investigaci\'on'' (AEI) and the EU ``Fondo Europeo de Desarrollo Regional'' (FEDER) through the project FPA2016-78645-P; and the Spanish MINECO through the ``Ram\'on y Cajal'' programme and through the Centro de Excelencia Severo Ochoa Program under grant SEV-2016-0597. MB also acknowledges support from the G\"oran Gustafsson foundation. 

\bibliographystyle{JHEP}
\bibliography{Oscillation_bib}

\end{document}